 \documentstyle[12pt]{article}

\def\yzero{\smash{\hbox{$y\kern-4pt\raise1pt\hbox{${}^\circ$}$}}}

\def\a{\alpha}
\def\e{\epsilon}
\def\te{\tilde{\epsilon}} 
\def\b{\beta}
\newcommand{\bmat}{\left(\begin{array}}
\newcommand{\emat}{\end{array}\right)}

\def\d{\delta}
\def\t{\times}

\def\la{\lambda}

\def\-{\hphantom{-}}
\def\ov{\overline}
\def\s2{\frac{1}{\sqrt2}}

\def\beq{\begin{equation}}
\def\eeq{\end{equation}}
\def\beqa{\begin{eqnarray}}
\def\eeqa{\end{eqnarray}}
\def\ba{\begin{array}}
\def\ea{\end{array}}

\def\IF{\relax{\rm I\kern-.18em F}}
\def\II{\relax{\rm I\kern-.18em I}}
\def\IP{\relax{\rm I\kern-.18em P}}
\def\IC{\relax\hbox{\kern.25em$\inbar\kern-.3em{\rm C}$}}
\def\IR{\relax{\rm I\kern-.18em R}}

\def\ti{\tilde}
\def\wt{\widetilde}

\def\cp{{\cal P}}

\def\Dsl{\,\raise.15ex\hbox{/}\mkern-13.5mu D} 
\def\IZ{Z\kern-.4em  Z}

 \def\cp#1{\relax\ifmmode {\IP\kern-2pt{}_{#1}}\else $\IP\kern-2pt{}_{#1}$\=fi}
\newcommand{\drawsquare}[2]{\hbox{%
\rule{#2pt}{#1pt}\hskip-#2pt
\rule{#1pt}{#2pt}\hskip-#1pt
\rule[#1pt]{#1pt}{#2pt}}\rule[#1pt]{#2pt}{#2pt}\hskip-#2pt
\rule{#2pt}{#1pt}}

\newcommand{\fund}{\raisebox{-.5pt}{\drawsquare{6.5}{0.4}}}
\newcommand{\Ysymm}{\raisebox{-.5pt}{\drawsquare{6.5}{0.4}}\hskip-0.4pt%
        \raisebox{-.5pt}{\drawsquare{6.5}{0.4}}}
\newcommand{\Yasymm}{\raisebox{-3.5pt}{\drawsquare{6.5}{0.4}}\hskip-6.9pt%
        \raisebox{3pt}{\drawsquare{6.5}{0.4}}}
\newcommand{\antifund}{\overline{\fund}}

\catcode`\@=11
\newdimen\@rotdimen
\newbox\@rotbox

\def\@vspec#1{\special{ps:#1}}
\def\@rotstart#1{\@vspec{gsave currentpoint currentpoint translate
   #1 neg exch neg exch translate}}
\def\@rotfinish{\@vspec{currentpoint grestore moveto}}
\def\@rotr#1{\@rotdimen=\ht#1\advance\@rotdimen by\dp#1%
   \hbox to\@rotdimen{\hskip\ht#1\vbox to\wd#1{\@rotstart{90 rotate}%
   \box#1\vss}\hss}\@rotfinish}
\def\@rotl#1{\@rotdimen=\ht#1\advance\@rotdimen by\dp#1%
   \hbox to\@rotdimen{\vbox to\wd#1{\vskip\wd#1\@rotstart{270 rotate}%
   \box#1\vss}\hss}\@rotfinish}%
\def\@rotu#1{\@rotdimen=\ht#1\advance\@rotdimen by\dp#1%
   \hbox to\wd#1{\hskip\wd#1\vbox to\@rotdimen{\vskip\@rotdimen
   \@rotstart{-1 dup scale}\box#1\vss}\hss}\@rotfinish}%
\def\@rotf#1{\hbox to\wd#1{\hskip\wd#1\@rotstart{-1 1 scale}%
   \box#1\hss}\@rotfinish}%
\def\rotate{\@ifnextchar[{\@rotate}{\@rotate[l]}}
\def\@rotate[#1]#2{\setbox\@rotbox=\hbox{#2}\@nameuse{@rot#1}\@rotbox}

\catcode`\@=12

\begin{document}

\makeatletter
\@addtoreset{equation}{section} \makeatother
\renewcommand{\theequation}{\thesection.\arabic{equation}}
\pagestyle{empty}

\begin{center}
\Large{\bf MSSM GUT String Vacua, Split Supersymmetry and Fluxes }
\\[10mm]
{\large{\bf 
E. G. Floratos$^{(1, 2)} $
and 
C. Kokorelis$^{(1, 2)}$
}}
\\[4mm] 
\normalsize{\small{ 
$^1$ Institute for Nuclear $\&$ Particle Physics, N.C.S.R. Demokritos,GR-15310, Athens, Greece}
\\[-0.3em]\normalsize{
\em 
$^2$  
Nuclear and Particle Physics Sector, Univ. of Athens,
GR-15771 Athens, Greece}}
\\[3mm]
\small{\bf Abstract}\\[1mm]
\end{center}
\begin{center}
\begin{minipage}[h]{14.5cm}
{\small We show that previous proposals to accommodate the MSSM with 
string theory N=0 non-supersymmetric compactifications coming from 
intersecting D6-branes
may be made fully consistent with the cancellation of RR tadpoles. In this respect 
we present 
the first examples of non-supersymmetric 
string Pati-Salam model vacua with starting observable gauge 
group $SU(4)_c \t SU(2)_L \t SU(2)_R$ (SU(2) from Sp(2)'s) that 
accommodate the spectrum  of the 3 generation MSSM 
with a gauged baryon number with
all extra exotics (either chiral or non-chiral)
becoming massive and all MSSM Yukawas realized. These constructions include models with
sin$^2(\theta_W) = \ 3/8$ (SU(5) type) and can have 1, 2  or  4 pairs of higgsinos depending 
on the $\#$ of tilted tori.
We work within four dimensional compactifications of IIA theory on 
toroidal orientifolds without (and with) fluxes.
The MSSM spectrum (together with right handed neutrinos)
is realized in 
the intersections of the visible sector that may contain D6-branes whose 
intersections 
share the same N=1 supersymmetry. The N=1 supersymmetry of the visible sector
is broken by an extra supersymmetry messenger breaking sector that preserves a 
different N$^{\prime}$=1 susy, exhibiting the first examples of stringy 
gauge mediated models.
Due to the high scale 
of the models, these models are also the first realistic examples of 
carriers of stringy  split supersymmetry
exhibiting universal
slepton/squark masses, massive string scale gauginos,  
{\em unification of SU(3), SU(2) gauge couplings at} $2.04 \times 10^{16}$  GeV, a stable proton 
and the appearance of a landscape split SM with chiral fermions and only Higgsinos 
below the 
scale of susy breaking; the LSP neutralino candidate could also be only Higgsino or Higgsino-Wino mixture. 
We also add RR, NS and metric fluxes 
as every intersecting D-brane model without fluxes can be accommodated in the 
presence of fluxes.
The addition of metric fluxes in the toroidal lattice also stabilizes the expected real parts of all in AdS 
closed string moduli (modulo D-term effects),  leaving unfixed only the imaginary parts of K\"ahler moduli. }

\end{minipage}                 
\end{center}

\newpage
\setcounter{page}{1}
\pagestyle{plain}
\renewcommand{\thefootnote}{\arabic{footnote}}
\setcounter{footnote}{0}

\section{Introduction}

Maybe the most serious problem of string theory nowadays is to derive a 
model of particle physics which will be as close as possible to the Standard 
Model at low energies and which not only manage to fix all its free parameters, 
its moduli, but it will also get rid of all its extra exotics (chiral or non-chiral)-- by making them massive through appropriate Yukawa couplings or some other mechanism --  which always have 
been a problem in model building attempts. The obvious next step to such an goal 
is to calculate specific phenomenological quantities that could make some definite
predictions for present and future experiments. 
In this respect recent model building attempts - without the presence 
of background fluxes - 
have been focused in the
construction of N=1 supersymmetric \cite{cve, cve1, bluo1, duti} and 
non-supersymmetric models \cite{imr, kokos1, kokos2a, kokos2b, kokosusy, D51, D52} and the use of
intersecting branes [See also \cite{ov}, \cite{ov1} --- for some semirealistic attempts 
in deriving the MSSM from another 
direction.]. These models make use of the fact that chiral fermions live in the
intersections of branes that intersect at angles or in other cases they make use
of the T-dual formulation of models with D9-branes and 
magnetic fluxes \cite{sa,bi, bitre}. Some comments are in order. \newline
In all recent string N=1 supersymmetric models \cite{cve,cve1,bluo1,ott1,she,alda,hone,ana}- where NSNS and RR tadpoles 
cancel - the localization 
of MSSM (we mean the usual multiplet context MSSM with one or more 
Higgsino $H_u$, $H_d$ multiplets and also right handed neutrinos)
 is accompanied by an unwanted  problematic 
large number of non-chiral or chiral open string exotics that survive massless to low 
energies \footnote{We also note that N=1 model building based on the by now
old heterotic string approach suffers from problems like 
proton decay and unfixed moduli parameters. See for example \cite{lebe}.}. 
These states could be coming from either the 
adjoint gauge 
multiplet (chiral ones) or from sectors (non-chiral ones) formed between in brane 
intersections where the participating branes are  
parallel in some 
compact direction. It is also possible that adjoint matter is formed from 
open string states that are accommodated in the intersection of a brane with its 
orbifold images. This is the case of $Z_N$ or $Z_N \times Z_M$ orientifolds of IIA 
compactifications with intersecting D6-branes. See for example \cite{bluo1}, \cite{orbiko}.

Following these developments during the era of development of 
intersecting brane models the initial expectation
was that possibly IIB string backgrounds 
in the presence of NS and RR fields \cite{blt, cu} will allow for more flexibility into 
the spectrum, such that the extra exotics will disappear from the spectrum 
or alternatively the 
present formulation of N=1 models with intersecting D6-branes will manage somehow 
to find a vacuum that have less \cite{cve1} or even no massless exotics at the end. 
Needless to say that in model building with or without fluxes surviving massless 
exotics to low energies are always present either as a part of 
a chiral or a non-chiral set of new exotic 
particles [for some examples see \cite{ms, she}.].

On the other hand hand non-supersymmetric models (NSM) from intersecting 
branes (without 
fluxes) \cite{imr, kokos1, kokos2a, kokos2b, kokosusy} in toroidal orientifold 
compactifications \cite{lustr} of type IIA theory [see also \cite{D51, D52} for the 
generation 
of the same SM configurations using D5-branes in different IIB backgrounds], 
have some successes as it become 
possible to derive for the first time vacua which have only the SM at low 
energies and with no extra chiral exotics - and with proton stability - without 
the presence 
of any additional chiral fermions at low energies. Hence the original four stack non-susy
SM \cite{imr} string vacua - that have no supersymmetry preserved at any intersection -
 have been extended \footnote{By spreading the SM particle context to different intersections.} to five and six stack SM string vacua that have one and two intersections
preserving a supersymmetry \cite{kokos2a, kokos2b} respectively \footnote{These models predict
the existence of the chiral spectrum of the SM in addition to only one type of supersymmetric particles, namely the susy 
partners of $\nu_R$'s the sneutrinos.}.
[For one way to get rid in these 
vacua of 
non-chiral fermions in any representation see e.g. \cite{ai}; one could also
try to get rid of only the adjoint matter but using fractional D-branes in 4D IIB 
chiral compactifications \cite{duti}].
These models follow a bottom-up approach as they embed the SM local configuration
to a string compactification and thus they are different to  
the top-bottom MSSM embedding approach of heterotic string compactifications.
We also mention at this 
point the construction of non-susy Pati-Salam vacua with only the SM at low
energy \cite{kokos1}, where all the accompanying beyond the SM extra chiral exotics 
become massive through the use 
of non-renormalizable mass couplings \footnote{
One important constraint that is derived in these models - proton is 
stable - is 
that the masses of the 
extra chiral exotics are greater than the electroweak scale only if the string scale 
is low and below 1.2 TeV. However the latter is unlikely to happen- the string 
scale is close to the Planck scale - as on these models  
the intersecting D6-branes 
wrap the whole of the toroidal orientifold space and there are no transverse 
dimensions to the 
D6-branes that could become large such that the string scale could be as lowered to
below 1.2 GeV. Some other possibilities which make these models consistent 
with a high scale will be explored elsewhere.}. In this work, we will see 
an identical effect - which is
consistent with a high scale - and
makes all exotics massive and leaves only the MSSM N=1 context at low energies 
[using the same orientifold backgrounds \cite{lustr} without fluxes]. 

In parallel with the development of non-susy models with only the SM at low
energy, a different direction was initiated in \cite{cre},
and further explored 
in \cite{lo1}, which localized the MSSM on a 4D four stack toroidal 
orientifold IIA
vacuum (no fluxes present). 
In this case, even though an anomaly free configuration was found
that locally corresponded to the MSSM N=1 multiplet spectrum nevertheless RR tadpoles
could not be satisfied. In \cite{kokosusy} we generalized the considerations of 
\cite{cre} to the maximal five and six stack locally supersymmetric MSSM without 
fluxes and also
considered the introduction of tilted tori on four, five and six stack MSSM local 
configurations \footnote{By brane recombination the five, six stack local
 MSSM models of
\cite{kokosusy} flow to the four stack MSSM local configurations of \cite{cre, lo1}.}. 
Furthermore in \cite{cre} it was argued 
that since the MSSM local configuration was anomaly free in order to satisfy RR tadpoles either an  
extra anomaly free sector was needed or NS-NS background fluxes that however should add no net chiral
content. Additionally, in 
\cite{kokosusy} we argued \footnote{The 5- and 6-stack MSSM local configurations 
of \cite{kokosusy} flow - under brane recombination of the U(1)'s - to the 4-stack models 
of \cite{cre,lo1}.} 
that the extra sector should play the role of the
supersymmetry breaking sector of gauge mediation \cite{gm} and the states of the
extra RR cancelling sector should be vector-like. We should mention that the same 
local configuration
describing the MSSM \cite{cre}, \cite{lo1} has been also used for the 
studies of unification of 
gauge couplings \cite{blu} and calculation of soft 
terms without fluxes \cite{kane2} in intersecting brane models.

The purpose of the present paper is twofold.
Initially we present a new method of cancelling RR 
tadpoles in N=0 (non-supersymmetric) toroidal orientifold models by adding a 
vector-like $N^{\prime}=1$ 
supersymmetric sector to N=1 local MSSM 
configurations, leaving as the only net chiral context of the theory the  
multiplet spectrum of the N=1 Standard Model. The vector-like sector respects
a different N=1 supersymmetry than the one respected by the sector that localizes the
MSSM. Hence overall SUSY is broken and the models are non-supersymmetric. Furthermore,
since we show that all extra beyond the MSSM matter either chiral or non-chiral becomes massive  
this is the first appearance of a realistic string model that finds the N=1 MSSM 
particle 
content inside a model where SUSY is already broken \footnote{At present there is no N=1
 supersymmetric models which manage in some way to get rid of its extra exotics chiral and/or 
non-chiral ones that accompany the MSSM spectrum.
}[overall the model is 
non-supersymmetric]. At present there is no N=1 model that can localize the 
MSSM matter context and be able to get rid of its massless exotics chiral and/or 
non-chiral ones.
\newline
 We apply this method to toroidal orientifolds by finding the most general solution
to RR tadpoles that localize the MSSM multiplet spectrum inside a Pati-Salam
$SU(4)_c \t Sp(2)_L \t Sp(2)_R$ construction. {\em What we call the MSSM 
is the usual MSSM chiral multiplet spectrum with right handed neutrinos - with gauged baryon number and 
hence proton stability - and either 
one (1), two (2) or 
four (4)
pairs of Higgsinos }$H_u$, $H_d$ ; the latter choices depending on the number of tilted tori.\newline 
 Furthermore we will show that 
all the extra beyond the MSSM matter - all the states of the extra vector-like 
messenger sector will get masses from Yukawa 
couplings.  Adjoint matter will get masses
from the introduction of Scherk-Schwarz breaking. 
On the other hand - following recent important developments on moduli 
stabilization either in N=1 supersymmetric AdS vacua with NS/R fluxes \cite{DeWolfe} 
and 
N=1 supersymmetric vacua with NS/R and metric fluxes \cite{cfi} recently, where all the 
real parts of the 
moduli could get fixed and only some combinations of axions are being left 
unfixed - we examine the issue of moduli stabilization by adding metric fluxes in 
the present 
models. We
recall that in non-susy compactifications \cite{imr, kokos1, kokos2a, kokos2b} from 
toroidal orientifolds 
only complex moduli 
could get fixed through the use of supersymmetry conditions on intersections; see for
example \cite{kokos1}. {\em Moreover the presented Pati-Salam models - when Scherk-Schwarz 
breaking is included -
have all the necessary ingredients to be the first realistic split susy \cite{split} 
models
from string theory \cite{split1}, \cite{split2}, \cite{orbiko}, \cite{split3} 
as the models exhibit partial unification of the strong SU(3) and weak SU(2) 
force
gauge couplings at the famous value $2.04 \times 10^{16}$ GeV as at the string scale only the MSSM particle content remains massless}.
    \newline
The structure of the present paper is as follows.
In section 2 we describe the local MSSM configurations as described in \cite{lo1}, 
\cite{kokosusy}
with zero and non-zero NS B-field respectively which did not satisfy at the time RR 
tadpoles \footnote{Even though in a revised version of \cite{kokosusy} a new RR tadpole canceling
model may be studied.}. 
In section 3 we present a class of RR tadpole solutions that describes the Pati-Salam embedding
$SU(4)_c \t Sp(2)_L \t Sp(2)_R$ type of models where all the additional exotics become 
massive and where the extra messenger - beyond the MSSM gauge group - is just two 
broken U(1)'s.  These RR tadpole solutions embed consistently the MSSM spectrum in the string 
compactification of four dimensional toroidal orientifolds \cite{lustr}.
We also show how all the chiral and non-chiral beyond the MSSM exotics become massive.
In section 4, 
the breaking of these models to its 
left-right counterpart symmetric models and the breaking to the MSSM is described also describing the way all
the chiral and non-chiral beyond the MSSM exotics become massive.
We also use complex structure moduli in Fayet-Iliopoulos terms 
to show that all sparticles/squarks get massive from the breaking of 
$N=1$, $N^{\prime} =1$ supersymmetries.
In section 5, the relation of the present constructions to split supersymmetry 
models and the satisfaction of all criteria for its application to strings set out in
\cite{split1, split2} is described. We also present a) 
constraints on the complex structure moduli derived from the requirement that the
Pati-Salam models contain an SU(5) type of sin$^2(\theta)$ at $M_s$. 
as well consequences for dark matter candidates.
In section 6 we describe the addition of R, NS and metric fluxes as 
described in \cite{vz}, \cite{cfi} and the fixing of most of the moduli in the Ads vacuum of the 
present models. 
We present our final comments and conclusions in section 7.

\section{MSSM and N=1 Supersymmetry in N=0 \\ Toroidal Orientifolds from 
Intersecting D6-branes}

\subsection{Preliminaries}

In string models from intersecting branes chiral fermions get 
localized in the intersection between branes. 
Lets us describe in some detail one such construction 
that involves four dimensional intersecting 
brane models coming from toroidal orientifolds of type IIA \cite{lustr}
theory. Consider IIA theory with D9-branes with fluxes compactified on a six dimensional 
torus, which is acted upon the worldsheet 
parity $\Omega$. Performing a T-duality on the 
$x^4, x^6, x^8$ compact directions the worldsheet parity action symmetry is mapped to 
$\Omega R$, where the action $R : z \rightarrow {\bar z}$
and the D9-branes with magnetic flux ${\em F}$ get mapped to
D6-branes intersecting at angles $\theta = tan^{-1}\frac{m R_2}{n R_1}$ , where 
$(n_a, m_a)^i$ the number of times the D6-brane $a$ is wrapping a one cycle on
 the i-th two torus. The $\Omega R$ action introduces an $\Omega R$ image of the
D6-brane that is wrapping the cycle $(n, -m)$. The introduction of discrete
B-field \cite{bfield} tilts the torus and maps the wrapping numbers according to the 
rule $(n, m) \rightarrow (n, m + (n/2))$. 
For this reason we introduce K-stacks of $N_a$, $a=1,...K$ of
D6-branes  wrapped on
$(n_a^i, m_a^i)$ cycle on the i-th two torus.
The net number of chiral fermions localized in the intersection between the branes
$a, b$ and $a, b^{\star}$ is then given respectively \footnote{The intersection number is the product of the homology 
classes of the D$6_a$-branes $\Pi_a$ and their orientifold images
$\Pi_a^{\star}$ , where we define the homology classes of the three cycles 
$[a_i ]$, $[ b_i ]$ of the ith-torus as
\beq
[ \Pi_a ] = \prod_{i=1}^3 
(n_a^i [ a_i ] + m_a^i [ b_i ] ), \ \ [ \Pi_{a^{\star}} ] = \prod_{i=1}^3 
(n_a^i [ a_i ] - m_a^i [ b_i ] )
\eeq
}
\beqa
I_{ab} &=&  [ \Pi_a ] \cdot [ \Pi_b ]  \ = \ \prod_{i=1}^3 (n_a^i m_b^i - n_b^i m_a^i), \nonumber\\ 
I_{ab^{\star}} &=&  [ \Pi_a ] \cdot [ \Pi_b^{\star} ]  \ = \ -\prod_{i=1}^3 (n_a^i m_b^i + n_b^i m_a^i),
\eeqa
for open string states starting from brane a and ending on brane b. Such a 
state belongs to a bifundamental $(N_a, {\bar N}_b)$ of the gauge group, e.g 
a left handed quark. 
Chirality is defined by choosing an intersection sign, 
negative sign implies right handed particles. 
The spectrum of toroidal orientifolds could also accommodate 
fermions in the S + A representations \footnote{The sector denoted as 
$aa^{\prime}$ in table (\ref{matter}) also involves intersections of the a-brane 
with the orientifold O6 plane.} of the gauge group. A gauge group U(N) 
appears from sectors that involve open strings that start and end on the same 
stack of N coincident 
branes. If a brane is its own orientifold image then an Sp(2) gauge group can 
appear.

\begin{table}[htb] \footnotesize
\renewcommand{\arraystretch}{1.25}
\begin{center}
\begin{tabular}{|c|c|r|||||}
\hline\hline
\hspace{2cm} {\bf Sector} \hspace{1cm} &  Multiplicity
& {\bf Representation} \hspace{2cm} \\
\hline\hline
$aa$  &       & $U(N_a)$ vector multiplet \hspace{3.3cm} \\
       & 3  & Adj. chiral multiplets  \hspace{3.3cm} \\
\hline\hline
$a(b)$   & ${\tilde I}_{ab} $ & $(\fund_a,\antifund_b)$ fermions  \hspace{3.3cm} \\
\hline\hline
$a(b')$ & ${\tilde I}_{ab^{\prime}}$ & $(\fund_a,\fund_b)$ fermions \hspace{3.3cm}  \\
\hline\hline
$a(a^{\prime})$ & 
$4 m_a^1 m_a^2 m_a^3 ( n_a^1 n_a^2 n_a^3 - 1) $ &
$\Ysymm\;\;$    fermions \hspace{3.3cm}   \\
  & $ 4 m_a^1 m_a^2 m_a^3 ( n_a^1 n_a^2 n_a^3 + 1)   $ &
$\Yasymm\;\;$   fermions  \hspace{3.3cm} \\
\hline\hline
\end{tabular}
\end{center}
\caption{\small General spectrum on D6-branes  at generic angles
(namely, not parallel to any O6-plane in all three tori ) in toroidal 
orientifolds. 
The models contain additional non-chiral pieces
in the $aa'$, $ab$, $ab'$ sectors with zero intersection, if
the relevant branes have a parallel direction. The latter could become 
massive in principle
when SS breaking is included. 
\label{matter} }
\end{table}

The above rules of determining the gauge group and chiral spectrum are not enough
when deriving an extension of the Standard Model from a string compactification.
Another consistency condition that may be satisfied by the D6-branes wrapping 
the compact space are the RR tadpole
cancellation conditions for compactifications which is the cancellation of the RR 
charge in homology 
\beq
\sum_a N_a [ \Pi_a ] \ + \  \sum_{a} N_a [ \Pi_{a{\star}} ]\ =  \ 32 [ \Pi_{O_6} ]
\eeq
for the D-branes and the their orientifold images and their O-planes. 
For four dimensional compactifications
of type IIA on toroidal 
orientifolds with D6-branes intersecting at angles they are given by
\beqa
\sum_a \ \ N_a \ n_a^1 \  n_a^2 \ n_a^3  &=& 16,\nonumber\\ 
\sum_a  N_a \ n_a^1 \  m_a^2 \ m_a^3  &=& 0, \nonumber\\
\sum_a  N_a \ m_a^1 \  n_a^2 \ m_a^3  &=& 0, \nonumber\\
\sum_a  N_a \ m_a^1 \  m_a^2 \ n_a^3  &=& 0
\label{RR}
\eeqa
 
Another consistency condition which affects the chiral spectrum is the 
existence of some amount of supersymmetries carried out by the D-branes.  

The notation for supersymmetries shared by the intersections and the orientifold 
planes is as follows. 
A pair of intersecting D6-branes that wraps across the $T^6$ factorizable 
tori and having angles $\theta_i$ across the $T^6$ , i=1,2,3 preserves
N=1 supersymmetry  if
\beqa
\ \pm \ \theta_1 \ \pm \ \theta_2 \ \pm  \ \theta_3 &=& 0
\eeqa
for some choice of the signs, where the angles $\theta_i$ are the relative 
angles between a pair of branes across the three 2-tori.
We distinguish the different susys shared by the branes and the orientifold planes 
by the sign choices \footnote{We follow the relevant discussion in \cite{Iba}.}
\beqa
r_1 &=& (\frac{1}{2})(-++-)\nonumber\\
r_2 &=& (\frac{1}{2})(+-+-)\nonumber\\
r_3 &=& (\frac{1}{2})(++--)\nonumber\\
r_4 &=& (\frac{1}{2})(----) 
\eeqa
which correspond respectively to the angles choices
\beqa
- \ \theta_1 \ + \ \theta_2 \ + \ \theta_3 &=& 0,\nonumber\\
+ \ \theta_1 \ - \ \theta_2 \ + \ \theta_3 &=& 0,\nonumber\\
+ \ \theta_1 \ + \  \theta_2 \ - \ \theta_3 &=& 0,\nonumber\\
+ \ \theta_1 \ + \ \theta_2 \ + \ \theta_3 &=& 0 \ .
\eeqa 
The masses of the 
lightest scalar states appearing in the NS sector 
in an intersection are 
\beqa
m^2 = \frac{M_s^2}{2}(- \ \theta_1 \ + \ \theta_2 \ + \ \theta_3 ), 
& m^2 = \frac{M_s^2}{2}( \ \theta_1 \ - \ \theta_2 \ + \ \theta_3 ), 
\nonumber\\
m^2 = \frac{M_s^2}{2}( \ \theta_1 \ + \ \theta_2 \ - \ \theta_3 ),
& m^2 = \frac{M_s^2}{2}( 1- \frac{1}{2}
(\ \theta_1 \ + \ \theta_2 \ + \ \theta_3))\ .
\label{mainter}
\eeqa

At present there are three ways to embed the Standard Model (or the MSSM) gauge 
group and spectrum into a unitary or symplectic configuration. \newline
They are classified according to the "observable" gauge group that appears at the string scale and they are grouped 
into three classes
of models, namely the first one (\ref{fir1}), (\ref{fir2}); the second class 
(\ref{fir3}), (\ref{fir4}) and the third class (\ref{third}) which read 
\newline
$\bullet$ \beq
U(3)_c \times U(2)_L \times U(1)^n ; \ \ n = 2, 3, 4 
\label{fir1}
\eeq
\beq
U(4)_c \times U(2)_L \times U(2)_R \times U(1)^m ;\ \ m \neq \ 0  \ ,
\label{fir2}
\eeq
and \newline
$\bullet \bullet$
\beq
U(3)_c  \times Sp(2)  \times U(1)  \times U(1) \ ,
\label{fir3}
\eeq
\beq
U(4)_c  \times Sp(2)  \times Sp(2) 
\label{fir4}
\eeq
and \newline
$\bullet \bullet  \bullet $
\beq
U(4)_c  \times Sp(2)_L  \times U(2)_R, \ or  \ \ U(4)_c  \times U(2)_L  \times Sp(2)_R,
\label{third}
\eeq

The first class of three generation models, namely (\ref{fir1}), (\ref{fir2}),
 uses only 
bifundamental fields 
for the chiral field description of the open string spectrum and treats the weak
group $SU(2)_L$ as the one coming from the 
decomposition $U(2) \supset SU(2)_L \t U(1)$. \newline
Such constructions have been 
discussed in the context of non-supersymmetric constructions with the chiral 
spectrum of the SM at low energy 
in \cite{imr} (n=2) for four-stack models and  in \cite{kokos2a, kokos2b}(n = 3, 4) respectively for related constructions
that represent the maximal SM embedding in the five- and six-stack extensions of \cite{imr} 
respectively. 
On the other hand 
the Pati-Salam \cite{ps} embedding 
(\ref{fir2}) was discussed in the construction of non-supersymmetric GUTS in
\cite{kokos1}. The SM embedding \cite{imr,kokos2a, kokos2b, kokos1} follows a bottom-up 
approach in the sense that 
they embed anomaly free configurations that describe the SM chiral spectrum 
\cite{imr, kokos2a, kokos2b, kokos1} into overall N=0
non-supersymmetric string models \footnote{N=1 string models will not be mentioned explicitly 
in the following considerations but we will comment where necessary.  In general 
these models are not very appealing at the moment as 
they are full 
of extra beyond the MSSM exotics which survive massless to low energies. At present 
there is no know way to get rid of these exotics in any N=1 construction that involves 
D-branes intersecting or not.} with complete cancellation of RR tadpoles. Hence these are 
genuine string \footnote{See \cite{leo} for RG gauge group studies of non-string models in 
a D-brane inspired 
context.}
models \footnote{An embedding that localizes the chiral SM spectrum have been also found in 
compactifications   
of $Z_3$ \cite{z3} or $Z_3 \times Z_3$ \cite{orbiko} orientifolds of type IIA using
intersecting D6-branes. In this case the SM is made of mixtures from 
bifundamental and antisymmetric representations.} embedded in 
four dimensional toroidal orientifolds of type IIA theory with intersecting 
D6-branes \cite{lustr}. These models have a stable proton, as baryon number is a 
gauged symmetry and the corresponding global symmetry survives to low energy as the
corresponding gauge boson becomes massive, while in some of these models neutrinos get 
masses from 
quark condensates (QCs) \cite{imr}, \cite{kokos1}(for recent discussions on QCs 
see \cite{ki}).

In the second class of models, namely the constructions given by 
(\ref{fir3}),(\ref{fir4}),
where three generation non supersymmetric constructions are realized, 
the weak $SU(2)_L$ gauge symmetry is delivered 
within an $Sp(2) \equiv SU(2)$ factor. In this case, also the 
left-right construction utilizes the $SU(2)_R$ factor as an 
$Sp(2)_R$. \newline
Lets us note that while the 
non-supersymmetric (non-susy) models of the first class (\ref{fir1}), (\ref{fir2}), 
are useful in deriving the Standard Model
spectrum, non-susy models of the second class (\ref{fir3}), (\ref{fir4}), are useful 
to build the 
MSSM intersection numbers that are associated with the 
N=1 chiral
multiplet spectrum of the MSSM \cite{cre, lo1, kokosusy} in the context of a
non-supersymmetric string construction. 
Models of those types have been discussed in \cite{cre, lo1, kokosusy} and at present
they do not satisfy RR tadpoles. The purpose of his work is to bridge this literature 
gap and to show that the embedding of these models in a string construction is 
possible; so that model building will be elevated from the local \cite{cre,lo1,kokosusy}
to a string level. 
\newline
The third class of constructions seen in eqn. (\ref{third}), 
do not exist in the context
of non-supersymmetric string constructions from intersecting branes and the 
embedding of the three generation Pati-Salam non-susy constructions - into a toroidal 
orientifold compactification of type IIA with 
intersecting D6-branes \cite{lustr}  - which localizes the MSSM chiral 
multiplet spectrum may be described in \cite{kokosusy}.\newline
We mention that N=1 constructions from intersecting D6-branes where part of the 
gauge group
is as in eqn. (\ref{fir4}) have been considered in \cite{cve10} in the presence of only RR and NS fluxes. 
Unfortunately, at present there is no either compactification or known way to get rid
from the rich varieties of extra chiral fields that appear in all N=1 4D 
supersymmetric models from intersecting D6-branes with or without general fluxes. 
Also other MSSM-like N=1 constructions with the gauge group as in (\ref{fir4}) have been considered
\cite{alda} in the context of Gepner-type IIB orientifold compactifications.

\begin{table}
[htb]\footnotesize
\renewcommand{\arraystretch}{1.5}
\begin{center}
\begin{tabular}{||c||c|c|c||} 
\hline
\hline
$\ N_i$ & $(n_i^1, \ m_i^1)$ & $(n_i^2, \ m_i^2)$ & $(n_i^3, \ m_i^3)$\\
\hline\hline
 $N_a=3$ & $(1,\ 0)$  &
$(1/\rho, \  3 \rho )$ & $(1/\rho, \ -3\rho )$  \\
\hline
$N_b=1$  & $(0, 1)$ & $(1,\ 0)$ & 
$(0, \ -1)$ \\
\hline
$N_c=1$ & $(0, \ 1)$ & $(0,\  -1)$  & 
$(1, \ 0)$ \\\hline
$N_a=1$ & $(1,\ 0)$  &
$(1/\rho, \  3 \rho)$ & $(1/\rho, \ -3\rho )$  \\
\hline
\end{tabular}
\end{center}
\caption{\small D6-brane wrapping numbers \cite{cre, lo1} with orthogonal tori 
that gives rise to a three generation N=1 MSSM 
spectrum via the Pati-Salam $SU(4)_c \t SU(2)_L \t SU(2)_R$ gauge group. 
The general solution to the intersection numbers is parametrized by the a parameter 
$\rho = 1, 1/3.$
\label{spe}}          
\end{table}

\begin{table}
[htb]\footnotesize
\renewcommand{\arraystretch}{1.5}
\begin{center}
\begin{tabular}{||c||c|c|c||} 
\hline
\hline
$\ N_i$ & $(n_i^1, \ m_i^1)$ & $(n_i^2, \ m_i^2)$ & $(n_i^3, \ m_i^3)$\\
\hline\hline
 $N_a=3$ & $(1,\ 0)$  &
$(1/\rho, \  3 \rho \epsilon \beta_1)$ & $(1/\rho, \ -3\rho {\tilde \epsilon} 
\beta_2)$  \\
\hline
$N_b=1 
$  & $(0, \ \epsilon {\tilde \epsilon})$ & $(1/\beta_1,\ 0)$ & 
$(0, \ -{\tilde \epsilon})$ \\
\hline
$N_c=1$ & $(0, \ \epsilon)$ & $(0,\  -\epsilon)$  & 
$({\tilde \epsilon}/\beta_2, \ 0)$ \\\hline
$N_a=1$ & $(1,\ 0)$  &
$(1/\rho, \  3 \rho \epsilon \beta_1)$ & $(1/\rho, \ -3\rho {\tilde \epsilon} 
\beta_2)$  \\
\hline
\end{tabular}
\end{center}
\caption{\small General D6-brane wrapping numbers \cite{kokosusy} that gives rise to a 
three generation N=1 MSSM 
spectrum via the Pati-Salam $SU(4)_c \t SU(2)_L \t SU(2)_R$ 
gauge group when   $\e = {\tilde \e} = + 1,\ \e = -{\tilde \e}=+1 $.
The general solutions to the intersection numbers 
are parametrized by two phases $\e = \pm 1$, ${\tilde \e} \pm 1 $, the NS background
on the second and third tori respectively $\beta_1$, $\beta_2$ and a parameter 
$\rho = 1, 1/3.$
\label{spe1}}          
\end{table}

Explicitly, the first attempt \footnote{In N=1 models the MSSM chiral 
spectrum appears as part of the 
spectrum but such attempts produce exotics surviving massless to low energies. 
} to localize the MSSM within string theory 
D-brane non-supersymmetric 
models has been initiated in \cite{cre, lo1}
where only the 
intersection numbers - in the absence of a discrete NS B-flux - 
that localize the 
spectrum of the MSSM were provided (in a four stack construction) as they are seen in 
table (\ref{spe}). These wrappings give rise to an $U(3) \t Sp(2) \t Sp(2) \t U(1)$
gauge group from the intersecting D6-brane stacks a, b, c, d respectively.  
Also the choice of parallel stacks in table (\ref{spe})
suggests the gauge symmetry enhancement $U(3)_a \t U(1)_d \rightarrow U(4)$
by turning on suitable vevs for the adjoint multiplets of the model.
However this choice do not satisfy RR tadpoles in the 
simplest of the string constructions that is
toroidal orientifolds $T^6/{\Omega R}$ of IIA \cite{lustr}. 
\newline
The four stack intersection number 
configurations of table (\ref{spe}) have been generalized in the presence of the 
discrete NS B-flux in
\cite{kokosusy} hence suggesting  
the maximal five and six stack extensions \footnote{Represent deformations of the 
models
of \cite{cre, lo1}.} of table (\ref{spe}), again in the absence of any solutions to RR 
tadpole cancellation conditions; they are shown in table (\ref{spe1}). The latter solutions 
depend on several parameters, the NS-B-fields $\beta_1$, $\beta_2$ of the second and third tori; 
the phases $\e$, ${\ti \e}$ that can take the values $\pm 1$ and the parameter $\rho =1, 1/3$. 

Notice that {\em the solutions of table (\ref{spe1}) for the special values of the parameters
\beq
\beta_1 = \beta_2 = 1, \ \ \e ={\tilde \e} = 1, \ \  \rho =\frac{1}{3}
\label{val}
\eeq
with all the tori orthogonal, 
as they appear in the RR tadpole solution of the Pati-Salam models of table (\ref{newska}), 
reproduce in the top part the wrappings numbers of the MSSM Pati-Salam embedding 
of \cite{cre,lo1} and 
also for the values  of the parameters (\ref{val}) the observable 
Pati-Salam sector
embedding of the
MSSM intersection numbers of the IIB MSSM-like fluxed embedded model  
of \cite{ms} }. \newline
Explicitly, 
 the starting gauge group for the parameter values (\ref{val}),
is an 
\beq
U(4)_c \t Sp(2)_L \t Sp(2)_R
\eeq
where the first stack of D6-branes gives rise to a $U(4)_c$ factor,
the second stack to an $SP(2)_L \equiv SU(2)_L$ gauge group  and the third 
stack to an $SP(2)_R \equiv SU(2)_R$ gauge group, 
since the branes b, c are invariant under the 
$\Omega R $ action.
 The spectrum 
associated with the parameters (\ref{val}) is seen in table (\ref{cc1}). 
\begin{table}
[htb] \footnotesize
\renewcommand{\arraystretch}{1}
\begin{center}
\begin{tabular}{||c|c|c|c|c||}
\hline
Matter & $SU(4) \t Sp(2)_L \t Sp(2)_R$ & $I_{ij}$'s & $Q_a$ &   SYSY \\\hline
${F}_{L}$ & $3(4,\ 2, \ 1;\ 1, \ 1)$ & $(ab)$ &  $1$  &  $r_4$\\\hline
${\bar F}_{R}$ & $3({\bar 4},\ 1,\   2;\ 1, \ 1)$ & $(ac)$ &  $-1$  & $r_4$\\\hline
${\bar h}$ & $\frac{1}{\beta_1 \cdot \beta_2}(1,\ 2,\ 2; \ 1,\ 1)$ & $ (bc)$ & $0$   
& $r_1, r_4$\\\hline\hline
\end{tabular}
\end{center}
\caption{\small Chiral spectrum of a 3-stack 
D6-brane Pati-Salam extension of the MSSM with three generations of chiral multiplets.
We have chosen  
 $\rho=1/3$, $\epsilon ={\tilde \e} =1$, $\beta_1 = \beta_2 = 1$ in table (\ref{spe}).
\label{cc1}}
\end{table}
By adjoint splitting of the U(4) factor we get an $U(3) \t U(1)_d$  - which 
could be identified with $U(1)_{B-L}$ - thus 
recovering
a left-right extension $SU(3) \t SU(2)_L \t SU(2)_R \t U(1)_{B-L}$. Subsequently 
by also considering arbitrary positions and Wilson lines for the c D6-brane (see also 
\cite{lo1}) $SU(2)_R \rightarrow U(1)_c$, the original Pati-Salam model of 
table (\ref{cc1}) gives rise to the MSSM spectrum of table 
(\ref{newtab1})(as the spectrum exhibits N=1 supersymmetry as we will comment below),
\begin{table}
[htb] \footnotesize
\renewcommand{\arraystretch}{1.2}
\begin{center}
\begin{tabular}{||c|c|c|c|c|c|c||}
\hline
Matter Fields & Representation & Intersection & $Q_a$ & $Q_c$ & $Q_d$ & Y
\\\hline
 $Q_L$ &  $3(3, 2)$ & $(ab), (ab*)$ & $1$ & $0$ & $0$ &  $1/6$ \\\hline
$U_R$ & $3({\bar 3}, 1)$ & $(ac)$ & $-1$ & $1$ & $0$ &  $-2/3$ \\\hline    
 $D_R$ &   $3({\bar 3}, 1)$  &  $(a c^{\ast})$ &  $-1$ & $-1$ & $0$ &  $1/3$ \\\hline $L$ &   $3(1, 2)$  &  $({db}), (db^{\ast}) $ & $0$ & $0$ & $1$ & $-1/2$  \\\hline    
$N_R$ &   $3(1, 1)$  &  $(dc)$ &  $0$ & $1$ & $-1$ &  $0$  \\\hline    
$E_R$ &   $3(1, 1)$  &  $(dc^{\ast})$ & $0$ & $-1$ & $-1$ &  $1$  \\\hline 
$H_{d}$ & $\frac{1}{\beta_1 \cdot \beta_2}(1,2)$ & $(cb*)$ & $0$   & $ 1 $& $0$ &  $- 1/2$ \\\hline
$H_{u}$ & $\frac{1}{\beta_1 \cdot \beta_2}(1,2)$ & $ (cb)$ & $0$   & $ - 1 $& $0$ &  $ 1/2$ \\\hline
 \hline
\end{tabular}
\end{center}
\caption{\small Chiral spectrum of the four stack 
D6-brane N=1 Supersymmetric Standard Model with its 
$U(1)$ charges. The general form of the spectrum for 
non-trivial tilding along the second and third torus has been reproduced
from  \cite{kokosusy}. For $\beta_1 = \beta_2 =1$, it gives the local MSSM-like models 
of \cite{lo1}.
\label{newtab1}}
\end{table}
where only the hypercharge
\beq
U(1)^Y = \frac{1}{6}Q_a -\frac{1}{2}Q_c -\frac{1}{2}Q_d
\label{hypM}
\eeq
survives the Green-Schwarz mechanism massless to low energies

Lets us now describe some properties of tables (\ref{spe}), (\ref{spe1}), (\ref{cc1}), (\ref{newtab1}).

\begin{itemize}

\item Tables (\ref{spe}),  (\ref{spe1}): The intersection numbers are $I_{ab} = 3$, $I_{bc} = 3$, 
give rise to the usual multiplets of the Pati-Salam $G_{422}$ structure that 
accommodates three generations quarks and leptons into the following representations
\beqa
F_L &=& (4, 2, 1) =  q(3, {\bar 2}, \frac{1}{6}) + l(1, {\bar 2}, -\frac{1}{2}) 
\equiv (u, d, l)
\nonumber\\ 
{\bar F}_R &=& ({\bar 4}, 1, 2) = u^c ({\bar 3}, 1,  -\frac{2}{3}) +  
d^c ({\bar 3}, 1, \frac{1}{3}) + e^c (1, 1, 1) + N^c (1, 1, 0) \equiv (u^c, d^c, l^c)
\nonumber\\ 
\label{pss}
\eeqa
The quantum numbers on the right hand side of (\ref{pss}) correspond to the 
decomposition of $SU(4)_C \t SU(2)_L \t SU(2)_R$ under $SU(3)_C \t SU(2)_L \t 
U(1)_Y$; $l = (\nu, e)$ the lepton doublet of the SM and 
$l^c = (N^c, e^c)$ the right handed leptons.

\item Table (\ref{spe1}): Note that the brane b-wrappings  give rise to either 
the group Sp(2)$_b$, when the brane b is its own orientifold image as happens in the 
case $\e = {\ti \e} = +1$,  $\e = -{\ti \e} = +1$.

\item Tables (\ref{spe1}), (\ref{cc1}): The spectrum exhibits N=1 supersymmetry  
as long as the following susy condition holds \footnote{where $\chi_i = R_2^i /R_1^i$
the complex structure moduli in the i-th torus.} 
\beq
\beta_1  \chi_2 \ = \  \beta_2  \chi_3 \ .
\label{susy}
\eeq
Also notice that the branes separately, share some susy with the orientifold 
plane as seen in table (\ref{cc1}).

\item Tables (\ref{spe1}),(\ref{cc1}),(\ref{newtab1}) : The Higgs sector arises from open strings 
stretching between the branes b, c. 
The branes b, c are parallel in the first tori giving rise to a non-chiral sector
with N=2 supersymmetry.  
Note that the intersection number $I_{bc}$ vanishes thus the net chirality in this 
sector is 
zero. Hence we recover from this sector $1/({\b_1 \b_2})$ chiral multiplets
in the (2, 2) representation
of $SU(2)_L \t SU(2)_R$. When breaking the $SU(2)_R \rightarrow U(1)_c$ by 
adjoint breaking or else,
one (2, 2) multiplet will split into two N=1 multiplets (2, 1), and (2, -1) under
$SU(2)_b \times U(1)_c$ which are identified as the MSSM Higgs multiplets 
$H_u$, $H_d$. Hence after the breaking of the left-right symmetry, we will 
get  $1/({\b_1 \b_2})$ $H_u$'s and an equal number of $H_d$ N=1 multiplets, 
which gives us three (3) versions of the MSSM with 
one (1), two (2) and four (4) pairs of N=1 Higgs multiplets $H_u$, $H_d$.   

\item Global symmetries of the SM could be identified with some of the 
U(1)'s appearing in table (\ref{newtab1}). 
Hence the baryon number may be identified 
as $3 B = Q_a$, lepton number $L = Q_d$. As a result of the couplings of the 
U(1)'s to RR fields since baryon number is a gauged symmetry and the corresponding 
gauge boson is getting massive; proton is  
stable perturbatively. Issues on proton stability in intersecting
branes can be found in \cite{klewi, afk, bur1, nathp, cr1}.

\end{itemize}

Summarizing  the number of Higgs multiplets which 
depends on the number of tilted tori reads
\beqa
\frac{1}{\b_1 \b_2 } &=&  \left\{   \ba{ll} 1, & \mbox{$\b_1 = \b_2 = 1$} \\
2, & \mbox{$( \b_1 , \b_2 )=(1, 1/2) \ or \  ( \b_1 , \b_2 )=(1/2, 1)$}\\
4, & \mbox{$\b_1 = \b_2 = 1/2$} \ .
\ea \right. 
\eeqa


\section{\Large{Embedding the MSSM in a Pati-Salam String Compactification}} 

The local construction \cite{lo1}, \cite{kokosusy} of tables 
(\ref{spe}), (\ref{spe1}), (\ref{cc1}), (\ref{newtab1}) respectively, represent the 
observable 
Pati-Salam (PS) extension of MSSM that in this section
may be embedded globally in a 4D toroidal IIA orientifold string compactification since RR tadpoles will be shown 
to be cancelled appropriately.
At this point we introduce a new way to cancel RR tadpoles.

$\bullet$ The observable sector ${\cal O}$ made of a, b, c, d D6-branes seen in 
table (\ref{cc1}) is made from
intersections that localize 
the MSSM chiral multiplets and where all intersections respects the same N=1 
supersymmetry. We now cancel the RR tadpoles (\ref{RR}) by the addition of an extra sector 
that 
respects a $N^{\prime}=1$ supersymmetry and is made from the branes $h_1, h_2, h_3, h_4, h_5$  and which it is different than the N=1 respected 
by the "observable" Pati-Salam MSSM sector. 

$\bullet$ The wrapping numbers of the extra sector can be seen in 
table (\ref{ska}). 
Observe that a number of extra branes $h_1,h_2, .., h_5 $ have to be added to 
cancel RR tadpoles.
From these branes, only the $h_3$, $h_4$ one's have a non-zero intersection number
with the observable sector branes 
on all three tori and thus have extra net chiral exotic fermions localized in their 
intersection. 
Also we notice that since all intersections preserve either a  
$N=1$ or a $N^{\prime}=1$ supersymmetry, each fermion will also 
accompanied by its massless
boson part of the chiral multiplet. 
Non-chiral matter from sectors where the branes are parallel in some tori may 
be made 
massive as we discuss 
shortly. 
These RR tadpole solutions generate a PS embedding of the MSSM which 
may be studied in detail in the next section.


\subsection{Three generation Pati-Salam models with a Messenger Sector}

We will now derive unique PS string vacua where all
the extra exotics become massive.
In the wrapping numbers of table (\ref{spe1}) we set $\e = \te$.
The solution to the RR tadpoles may be seen in table (\ref{skaB}) where the 
$\rho$ parameter takes the value $1, 1/3$.

\begin{table}
[htb]\footnotesize
\renewcommand{\arraystretch}{2.3}
\begin{center}
\begin{tabular}{||c||c|c|c||c|||||} 
\hline
\hline
$\ N_i$ & $(n_i^1, \ m_i^1)$ & $(n_i^2, \ m_i^2)$ & $(n_i^3, \ m_i^3)$& G.G. 
$->$ $\epsilon =+1$  \\
\hline\hline
 $N_a=4$ & $(1,\ 0)$  & $(3, \  \e \beta_1)$ & $(3, \  -{\e} \beta_2)$ 
& $U(4)$  \\
\hline
$N_b=1$  & $(0, \ 1)$ & $(1/\beta_1,\ 0)$ & $(0, \ -{\e})$ & $Sp(2)$  \\
\hline
$N_c=1$ & $(0, \ \e)$ & $(0,\  -\e)$  & $({\e}/\beta_2, \ 0)$ & $Sp(2)$ \\    
\hline\hline\hline\hline
$N_{h^1} = 36\beta_1 \beta_2$ & $(1,\ 0)$  &
$(-1/\beta_1, \  0)$ & $(1/\beta_2, \ 0)$&$Sp(2)^{36\beta_1 \beta_2}$  
 \\\hline
$N_{h^2}=4\beta_1 \beta_2 $  & $(1, \ 0)$ & $(0,\ {\e})$ & $(0, \ {\e})$ & 
$U(1)^{4 \beta_1 \beta_2}$ \\\hline
$N_{h^3}=1$ & $(0, \ 1)$ & $(1/\beta_1,\  0)$  & $(0, \ {\e})$ & $U(1)$ \\    
\hline
$N_{h^4}=1$ & $(0,\ {\epsilon})$ &  $(0,\  \epsilon )$  
  & $({\epsilon}/\b_2, \ 0)$ & $U(1)$  \\\hline
$N_{h^5}=16\beta_1 \beta_2$ & $(1,\ 0)$ &  $(1/\beta_1,\ 0 )$  
  & $(1/\beta_2, \ 0)$ 
& $Sp(2)^{16\beta_1 \beta_2}$\\\hline\hline
\end{tabular}
\end{center}
\caption{\small 
Solution to the RR tadpoles for toroidal 
orientifold models. The N=1 MSSM chiral spectrum arises as part of Pati-Salam models 
in the top part of the table from intersections between a, b, c, d branes.  
Messenger multiplet states respecting a $N^{\prime}=1$ supersymmetry 
arise from the intersections of the a, d branes with the 
$h_3$, $h_4$ branes. 
We have set $\rho=1/3$, $\epsilon ={\tilde \epsilon}$ in table (\ref{spe1}).
\label{ska}} 
\end{table}

Substituting on table (\ref{skaB}) the value $\rho =1/3$, we obtain in table (\ref{newska}) the 
much wanted 
solutions to the
RR tadpoles for the local construction of \cite{lo1}. The chiral spectrum can be seen in 
table (\ref{appa1}).
The observable gauge group under which chiral fermions exist is 
\beq
U(4) \times U(2)_b \times U(2)_c \times U(1)_{h_3}  \times U(1)_{h_4} 
\label{obse1}
\eeq
or equivalently
\beq
SU(4) \times U(1)_a \times SU(2)_b \times SU(2)_c \times U(1)_{h_3}  \times U(1)_{h_4} \ ,
\label{gaugepa}
\eeq 
where the SU(2)'s come from Sp(2)'s; the chiral spectrum can be seen in table (\ref{appa1}).
and the gauge group 
(G.G) transformations and charges are under the (\ref{obse1}) G.G.
In the top part of table (\ref{appa1}) we find the usual Pati-Salam part that embeds the MSSM chiral 
spectrum while the extra sector that cancels RR tadpoles is seen in the bottom part. 
\newline
The branes, namely $h_1, h_2, h_5$ have no net chiral particle context with
the rest of the branes and thus they constitute a "hidden" sector. Also the
rest of gauge group factors associated with the "hidden" branes $h_1$,
$h_2$, $h_5$ do not give rise to chiral particles as they have no
intersections with the Standard model particles and the rest of the branes.
As we will see later
the non-chiral matter arising from these branes is made massive by the
introduction
of Scherk-Schwarz deformations. 
Regarding the notation of the gauge groups related to the $h_1 , h_2 , h_5$ stacks we have 
considered that each one of these stacks is made from single stacks.  To this end, 
non-chiral states that are coming from $h_1 h_1, h_5 h_5, h_1 h_5$ intersections should be 
properly considered as corresponding to the intersections  
$h_1^j h_1^j$, $h_5^v h_5^v$, $h_1^j  h_5^v$, where $j=1, ..., 36 \b_1 \b_2$, 
$v = 1,.., 16 \b_1 \b_2$.

\begin{itemize}

\item {\em \large{Neutrino Masses}} 
 
\end{itemize}

The bi-doublet "h" Higgs multiplets of table (\ref{appa1}) may be used in the process of 
electroweak symmetry breaking as the relevant Yukawa read
\beq
F_L {\bar F}_R h = \upsilon (u u^c + \nu N) + {\tilde  \upsilon} 
(d d^c + e e^c) \ , \ h = diag(\upsilon , {\tilde  \upsilon} )
\eeq
giving masses to the up-quarks and neutrinos.

\begin{table}
[htb]\footnotesize
\renewcommand{\arraystretch}{2.3}
\begin{center}
\begin{tabular}{||c||c|c|c||c|||||} 
\hline
\hline
$\ N_i$ & $(n_i^1, \ m_i^1)$ & $(n_i^2, \ m_i^2)$ & $(n_i^3, \ m_i^3)$& G.G. 
$->$ $\epsilon =+1$  \\
\hline\hline
 $N_a=4$ & $(1,\ 0)$  & $(3, \  1)$ & $(3, \  -1)$ 
& $U(4)$  \\
\hline
$N_b=1$  & $(0, \ 1)$ & $(1,\ 0)$ & $(0, \ -1)$ & $Sp(2)$  \\
\hline
$N_c=1$ & $(0, \ 1)$ & $(0,\  -1)$  & $(1, \ 0)$ & $Sp(2)$ \\    
\hline\hline\hline\hline
$N_{h^1} = 36$ & $(1,\ 0)$  &
$(-1, \  0)$ & $(1, \ 0)$&$Sp(2)^{36}$  
 \\\hline
$N_{h^2}=4 $  & $(1, \ 0)$ & $(0,\ 1)$ & $(0, \ 1)$ & $U(1)^{4}$ \\\hline
$N_{h^3}=1$ & $(0, \ 1)$ & $(1,\  0)$  & $(0, \ 1)$ & $U(1)$ \\    
\hline
$N_{h^4}=1$ & $(0,\ 1)$ &  $(0,\ 1 )$  
  & $(1, \ 0)$ & $U(1)$  \\\hline
$N_{h^5}=16$ & $(1,\ 0)$ &  $(1,\ 0 )$  
  & $(1, \ 0)$ 
& $Sp(2)^{16}$\\\hline\hline
\end{tabular}
\end{center}
\caption{\small 
Solution to the RR tadpoles for the toroidal 
orientifold models of \cite{lo1}. 
The N=1 MSSM chiral spectrum arises as part of Pati-Salam models 
in the top part of the table from intersections between a, b, c, d branes.  
Messenger multiplet states respecting a $N^{\prime}=1$ supersymmetry 
arise from the intersections of the a, d branes with the 
$h_3$, $h_4$ branes. 
\label{newska}}          
\end{table}

\begin{table}
[htb] \footnotesize
\renewcommand{\arraystretch}{2.5}
\begin{center}
\begin{tabular}{||c|c|c|c|c|c|c|||||||}
\hline
Matter & Repr. & $I_{ij}$'s & $Q_a$ & $Q_{h^3}$ & $Q_{h^4}$ &  SYSY \\\hline
${F}_{L}$ & $3(4,\ 2, \ 1;\ 1, \ 1)$ & $(ab)$ &  $1$  & $0$ & $0$ &  $r_4$\\\hline
${\bar F}_{R}$ & $3({\bar 4},\ 1,\  2;\ 1, \ 1)$ & $(ac)$ &  $-1$  & $0$ & $0$ &  $r_4$\\\hline
${\bar h}$ & $\frac{1}{\beta_1 \cdot \beta_2}(1,\ 2,\ 2; \ 1,\ 1)$ & $ (bc)$ & $0$   
& $0$ & $0$ & $r_4 , r_1$\\\hline\hline\hline\hline
${q}_{R}^1$ & $3({\bar 4},\ 1, \ 1;\ 1, \ 1)$ & $(ah_3)$ &  $-1$  & $1$ & $0$ &  $r_1$\\\hline
${q}_{R}^2$ & $3({\bar 4},\ 1, \ 1;\ 1, \ 1)$ & $(a{h_3}{\star})$ &  $-1$  & $-1$ & $0$ &  $r_1$\\\hline
${q}_{L}^3$ & $3(4,\ 1, \ 1 ;\ 1, \ 1)$ & $(ah_4)$ &  $1$  & $0$ & $-1$ &  $r_1$\\\hline
${q}_{L}^4$ & $3(4,\ 1, \ 1;\ 1, \ 1)$ & $(a{h_4{\star}})$ &  $1$  & $0$ & $1$ &  $r_1$\\\hline
\end{tabular}
\end{center}
\caption{\small Chiral spectrum of the Pati-Salam
D6-brane N=1 Supersymmetric Standard Model. 
The bottom part, that  
includes the massive $N^{\prime}=1$ messenger fields that communicate supersymmetry breaking
to the visible N=1 sector, completes the missing RR canceling sector 
of \cite{cre, lo1}.  
Note that the notation L,R at the bottom of the table is  
related to chirality and $\rho=1/3$, $\epsilon ={\tilde \e} = 1$.
\label{appa1}}
\end{table}

\begin{itemize}

\item {\em {\large Gauge mediation}}

\end{itemize}

The extra sector that is needed to cancel RR tadpoles introduces
a vector-like sector that plays the role of the messenger supersymmetry breaking sector 
of the gauge mediated models \cite{gm} and transforms under the observable $O$ and the extra
$h= \{ h_1, ..., h_5 \}$ sectors. The messenger extra sector needed for the RR tadpole cancellation is anomaly free and
possess a different N=1 supersymmetry that the one respected by the top MSSM 
embedding. 
Lets us mention that the spectrum at the top of table (\ref{appa1}) for $\b_1 = \b_2 = 1$, 
recovers the observable local MSSM spectrum of \cite{cre, lo1} and also the ``observable'' 
sector of the 
MSSM spectrum of \cite{ms}. However, in our case the messenger exotic multiplets may become massive. \newline
The questions that at this point remain are : \newline
a)  in which way we can get rid off
the chiral fermions that appear in the messenger sector and which make 
vector-like pairs of N=1 multiplets with respect to the "observable" SM 
$SU(3) \t SU(2)_w \ U(1)_Y$  \newline b) how we can get rid off the
adjoint fermions and gauginos that appear from brane-brane sectors \newline
c) the non-chiral non-adjoint fermions that appear in intersections where the
different participating intersecting branes are parallel in at least 
one complex plane.


\subsection{{\em \Large{Mass couplings for chiral fermions - in the Messenger Sector}}}


Before discussing in this section the mass couplings of the extra chiral fermions that make 
vector-like 
exotics of table (\ref{appa1}),
let us examine the sector formed from open strings stretching between the 
branes $h_3$, $h_4$. It is easy to see that in this sector the intersection number $I_{h_3 h_4}$ vanishes 
and also that 
the N=2 supersymmetries preserved by this sector are the $r_2, r_3$.  
This implies - since also these is no gauged hypercharge for this multiplets - that this sector acts as the usual Higgs sector between the b, c 
branes and 
 the net number of chiral fermions in this sector is zero. Hence there are 
$1/(\b_1 \b_2)$ equal numbers (the intersection number across the remaining 
non-parallel tori ) of the N=1 massless multiplets
\beq
h_{34}^{(1)}= (1, 1, 1; 1, 1)_{(0;1, -1)}\ , \  
h_{34}^{(2)} = (1, 1, 1; 1, 1)_{(0;-1, 1)}\ ,
\eeq
where the branes are parallel across the first tori\footnote{Where the notation regards the gauge group
representations $SU(4) \times SU(2)_L \times SU(2)_R \times U(1)_{h_3} \times U(1)_{h_4}$
and charges
follow the table (\ref{appa1}).}. These fermions are singlets under the 
SM gauge group and thus can receive vevs. Observe the rather striking fact that 
these Higgs multiplets that 
help us to get rid of the extra vector-like exotics are equal in number to 
the usual 
MSSM multiplets H that appear
between the b, c branes and their number depends also on the shape of the tori of 
the compactification.  
\newline 
The chiral messenger fermions on the bottom part of the of table (\ref{appa1}) can obtain
Dirac masses from the following superpotential couplings
\beq
(q_R^1)_{(-1;1, 0)} \ (q_L^3)_{(1;0, -1)} \ 
 h_{34}^{(2)}  \ , \  \ (q_R^2)_{(-1;-1, 0)} \ 
(q_L^4)_{(1;0, 1)} \    h_{34}^{(1)}  \ 
\label{exp1}
\eeq 
The $h_{34}^{(1)}$,  $h_{34}^{(2)}$, represent flat directions in the
effective potential.
By - assuming - that their scalar components receive a vev 
all messenger multiplets become massive and disappear from the 
low energy spectrum. At this point - the models are at the string scale - the observable gauge group is that
of (\ref{gaugepa}), and the massless chiral multiplet fields are those of the
top part of table (\ref{appa1}) and also present there are non-chiral multiplets and the adjoint multiplets. Next we discuss how the non-chiral multiplets
become massive.

\subsection{\em \Large{Masses for non-chiral fermions}} 

{\em 
All non-chiral fermions in the present Pati-Salam models get massive
from the combined use of flat directions and Schwerk-Schwarz (SS) deformations.}

Non-chiral fermions (NCM) arise in sections that branes are parallel in at least 
one tori
direction in some 
complex plane. The introduction of SS deformations in directions parallel
to the directions that the intersecting D6-branes wrap
give masses to non-chiral fermions that have odd n-``electric'' wrapping numbers \cite{ai} in any representation.
In this work, we introduce SS deformations in all three complex tori.  
We will examine the case that all the tori are not tilted, $\beta_1 = \beta_2 = 1$.

NCM's in sectors 
$aa$, $aa*$, bb, $bb*$, cc, $cc*$, $h^i h^i$, $h^i h^{i*}$, i=1,2,3,4,5 get a 
mass from SS
deformations from n-wrappings that are odd in at least one tori, namely
the first, first, second, second, third, third, first, first, first, first, 
second, second, third, third, third, third ones respectively.

\begin{itemize}

\item {\large{\bf Masses for NCM's in messenger sector}}

\end{itemize}
There are also
NCM from the sectors $ah^1$, $ah^{1*}$, $ah^2$, $ah^{2*}$, 
$ah^5$, $ah^{5*}$, 
$bh^3$, $bh^{3*}$, $bh^4$, $bh^{4*}$, $ch^3$, $ch^{3*}$, 
$ch^4$, $ch^{4*}$, where the participating intersecting branes (PIB's) are 
parallel in the first tori;
  $bh^1$,  $bh^{1*}$,
 $bh^3$, $bh^{3*}$, $bh^5$, $bh^{5*}$,
$ch^2$, $ch^{2*}$,
$ch^4$, $ch^{4*}$, where the PIB's are 
parallel in the second tori;
 $bh^2$,   $bh^{2*}$,
$bh^3$, $bh^{3*}$,
$ch^1$, $ch^{1*}$,
$ch^4$, $ch^{4*}$,
$ch^5$, $ch^{5*}$, where the PIB's are 
parallel in the third tori. 
All these NCM's get masses -- 
from SS deformations as they have odd n's -- but the ones from the intersections  
$bh^2$,  $bh^4$, $ch^2$, 
$ch^3$ (and their orientifold images) that appear in table (\ref{skla12}).  The latter ones's could get masses
from tree level flat directions. 
 
\begin{table}
[htb] \footnotesize
\renewcommand{\arraystretch}{1.5}
\begin{center}
\begin{tabular}{||c|c||||}
\hline
$Intersection$ & $ States $  \\\hline
$b h^2$ & $\ba{l}
\Omega_1 \ : \ (1,2,1;1,1,1,1,1)_{(0;-1,0,0)} \\
\Omega_2 \ : \ (1,{\bar 2},1;1,1,1,1,1)_{(0;1,0,0)}
\ea$ \\\hline
$b h^4$ & $
\ba{l}
\Omega_3 \ : \ (1,2,1;1,1,1,1,1)_{(0;0, 0,-1)} \\
\Omega_4 \ : \ (1,{\bar 2},1;1,1,1,1,1)_{(0;0,0,1)}
\ea
$ \\\hline
$c h^2$ & $
\ba{l}
\Omega_5 \ : \ (1,1,2;1,1,1,1,1)_{(0;-1,0,0)} \\
\Omega_6 \ : \ (1,1,{\bar 2};1,1,1,1,1)_{(0;1,0,0)}
\ea
$ \\\hline
$c h^{3}$ & $\ba{l}
\Omega_7 \ : \ (1,1,2;1,1,1,1,1)_{(0;0,-1,0)} \\
\Omega_8 \ :  \ (1,1,{\bar 2};1,1,1,1,1)_{(0;0, 1,0)}
\ea$ \\\hline\hline
$b h^{2*}$ & $\ba{l}
\Omega_9 \ : \ (1,2,1;1,1,1,1,1)_{(0;-1,0,0)} \\
\Omega_{10} \ : \ (1,{\bar 2},1;1,1,1,1,1)_{(0;1,0,0)}
\ea$ \\\hline
$b h^{4*}$ & $
\ba{l}
\Omega_{11} \ : \ (1,2,1;1,1,1,1,1)_{(0;0,0,-1)} \\
\Omega_{12} \ : \ (1,{\bar 2},1;1,1,1,1,1)_{(0;0,0,1)}
\ea
$ \\\hline
$c h^{2*}$ & $
\ba{l}
\Omega_{13} \ : \ (1,1,2;1,1,1,1,1)_{(0;-1,0,0)} \\
\Omega_{14} \ : \ (1,1,{\bar 2};1,1,1,1,1)_{(0;1,0,0)}
\ea
$ \\\hline
$c h^{3*}$ & $\ba{l}
\Omega_{15} \ : \ (1,1,2;1,1,1,1,1)_{(0;0, 1,0)} \\
\Omega_{16} \ : \ (1,1,{\bar 2};1,1,1,1,1)_{(0;0, -1,0,)}
\ea$ \\\hline\hline
\end{tabular}
\end{center}
\caption{\small 
Non-chiral multiplet states from Pati-Salam models that also persist in their
-- by adjoint breaking -- L-R symmetric models of section (4). The bottom part of the table 
includes the orientifold images of the top part multiplet states.  
\label{skla12}}
\end{table}

The following allowed superpotential couplings
\beqa
W & \sim & k_1 \Omega_1 \ \Omega_4 \ h_{24}^{(2)}  \ + \ 
 k_2 \Omega_2 \ \Omega_3 \ h_{23}^{(1)} \ + \ 
 k_3 \Omega_5 \ \Omega_8 \ h_{24}^{(2)} \ + \ 
 k_4 \Omega_6 \ \Omega_7 \ h_{23}^{(1)}  \nonumber\\
&+ &k_5 \Omega_9 \ \Omega_{12} \ h_{24}^{(1)} \ + \ 
 k_6 \Omega_{10} \ \Omega_{11} \ h_{23}^{(2)} \ + \ 
 k_7 \Omega_{13} \ \Omega_{16} \ h_{23}^{(1)} \ + \ 
 k_8 \Omega_{14} \ \Omega_{15} \ h_{23}^{(2)} 
\label{stat1}
\eeqa
generate Dirac masses for the fermion pairs $\Omega_1  \Omega_4$, 
$\Omega_2  \Omega_3$, $\Omega_5 \ \Omega_8$, $ \Omega_6  \Omega_7$,
 $\Omega_9  \Omega_{12}$, $\Omega_{10}  \Omega_{11}$, $\Omega_{13}  \Omega_{16}$, 
$\Omega_{14}  \Omega_{15}$.
We have used the notation $h_{ij}^{(l)}$ where by $i$, $j$ we denote
the N=2 multiplet appearing in the intersection between the branes $h^i$ and $h^j$. By
the superscript $l$ when $l=1,\ l=2$ we denote the states associated with the 
positive, negative intersection numbers $I_{ij}$ respectively. Also $k_i$ are numerical 
coefficients that can be calculated by the use
of string amplitudes and give rise to Yukawa couplings for fermions in the 
effective theory.
Lets us consider for example the superpotential term $k_1 \Omega_1 \ \Omega_4 \ h_{24}^{(2)}$ involving the multiplets $h_{24}^{(l)}$ that respects a N=2 susy. The $\Omega_1$ represents a 
N=1 multiplet preserving the supersymmetry $r_2$;  $\Omega_4$ represents a 
N=2 multiplet preserving the supersymmetries $r_2, r_3$; 
$h_{24}^{(2)}$ is preserving the susys $r_2, r_4$;
they all share the common N=1 susy $r_2$.\newline    
Also all gauginos and also adjoint multiplets get massive as the SS deformations act in all tori and there are odd $n_i$ in all branes in at least one tori direction.
Hence, only the observable MSSM multiplets seen at the top of table (\ref{appa1}) remain massless at the string scale. 
\begin{itemize}
\item  {\em Breaking to the MSSM}
\end{itemize}

The initial U(4) gauge symmetry breaks to $SU(4) \times U(1)_a$. The $U(1)_a$ gauge symmetry 
is getting massive from its couplings to RR fields and subsequently the $SU(4) \times SU(2)_L \times SU(2)_R$ symmetry breaks with the help of the Higgs multiplets states - localized in the intersection $ac*$
\beq
({\bar 4}, 1, 2), \ \ (4, 1, 2)
\label{use1}
\eeq
to the $SU(3) \times SU(2)_L \times U(1_Y$. We notice that $SU(4)$ breaks \footnote{
See also \cite{kokos1} for the breaking of Pati-Salam models that have sextets in their spectrum;
also 2nd ref. of \cite{cve1} for the breaking 
of the PS MSSM-like models in the context of $Z_2 \t Z_2$ orientifolds 
with intersecting D6-branes, where however massless exotics
also survive to low energies.} to 
$SU(3) \times U(1)_{B-L}$ and $SU(2)_R$ to $I_{3R}$ by (\ref{use1}).
We also note that the non-chiral multiplet states of table (\ref{skla12}) also appear in the left-right models
of the next section - with the addition of 
non-chiral states states that are formed between the intersections of the d- U(1) brane and the other branes
- and they get masses from the same mechanism described in this section.


\section{Flow to three generation Left-Right Symmetric Models and the MSSM}

We have seen that three generation Pati-Salam (PS) models of the previous section 
can have 
all its chiral and non-chiral states (beyond the MSSM) made massive leaving only the usual
Pati-Salam spectrum at $M_s$ and the SM at low energies. In this section, 
the breaking of these PS  
 models to a three generation left-right $SU(3) \t SU(2)_L \t SU(2)_R $ (L-R)
symmetric model with all its exotics made massive 
and the subsequent breaking of the L-R to the SM is described.
\newline\newline
$\bullet$ {\em Pati-Salam adjoint breaking to left-right symmetric models}
\newline\newline
The L-R model can be derived from the Pati-Salam models of the previous section
by considering the adjoint breaking of the $U(4)_c \rightarrow U(3)_c \t U(1)_d$.
The spectrum of these models can be seen in table (\ref{taba1}) for 
$\e = {\ti \e} = 1$ and $\rho =1/3$ while 
the U(1)'s appearing in tables (\ref{taba1}), namely $Q_a$, $Q_d$ are the U(1)'s 
that are inside U(3), U(4) respectively.

After examining the Green-Schwarz couplings 
\beq
 n^J n^K m^I \int B_2^I \wedge F_a, \ \  
n_a n_a n_a \int C^o \wedge F_a \wedge F_a ,\  n^I m^J m^K \int C^I \wedge F \wedge F
\label{coup10}
\eeq we find
that the mixed U(1) gauge anomalies cancel since the following couplings 
of the RR fields to U(1)'s exist : 
$B_2^3 \wedge (3 \epsilon \beta_2)(3 F_a + F_d)$; 
$(3 \epsilon \beta_1)(3 F_a + F_d) \wedge B_2^2$.
We find that the surviving  massless
the Green-Schwarz mechanism U(1)'s are the  
\beq
U(1)^{(3)} = F^{h_3} - F^{h_4} \ , \ U(1)^{(4)} =  F^{h_3} + F^{h_4} 
\eeq
and the hypercharge 
\beq
Q_Y \equiv U(1)^{(1)} = (1/6) \ F_a  - (1/2)\ F_d 
\label{sda1}
\eeq
In table (\ref{taba1}) we can see the hypercharge assignment of the left-right symmetric models. [In appendix A there is a different hypercharge assigment for the left-right symmetric models which breaks 
to the SM at low energies and also possess massive chiral and non-chiral exotics.]. 
The $U(1)^{(3)}$, $U(1)^{(4)}$ may be broken by vevs of the gauge singlets 
$\langle {\tilde \phi}_R^1 \rangle $, $\langle {\tilde \phi}_L^3 \rangle $ respectively. \newline
$\bullet$ {\em Global symmetries}\newline
The global symmetries that exist in the   
Standard Model do exist in the left-right symmetric models and 
get identified in terms of the U(1) symmetries set out by the D6-brane 
configurations of table (\ref{taba1}).
Hence baryon number ($B$) is identified as $Q_a = 3 B$ and the lepton 
number ($L$) could practically identified as $Q_d = L$ as the 
$R$-antimatter multiplet states accommodate both the right leptons.
\begin{table}
[htb] \footnotesize
\renewcommand{\arraystretch}{1.2}
\begin{center}
\begin{tabular}{||c||c||c||c|c|c||c|c||c|c||}
\hline
Matter & Repr. & $I_{ij}$'s & $Q_a$ & $Q_d$ & $Q_{h^3}$ & $Q_{h^4}$ 
& Y &  ${\ti Y}$    &SYSY 
\\\hline
 $Q_L$ &  $3(3,\ 2, \ 1, \ 1; \ 1,\ 1)$ & $(ab), (ab*)$ & $1$ & $0$ & $0$ & $0$ & $1/6$ &  $1/6$ & $N=1$ \\\hline
${Q}_R$ & $3({\bar 3},\ 1,\ 2, \ 1; \ 1, \ 1)$ & $(ac), (ac*)$ & $-1$ &  $0$ & $0$ & $0$& $-1/6$ & $-1/6$ & $N=1$\\\hline     
$L$ &   $3(1,\ 2, \ 1, \ 1; \ 1,\ 1)$  &  $({db}), (db^{\ast}) $ & $0$ & $1$ &  $0$ & $0$ & $-1/2$ & $-1/2$ & $N=1$ \\\hline    
$R$ &   $3(1,\ 1,\ 2,\ 1;\ 1,\ 1)$  &  $(dc), (dc*)$ &  $0$ & $-1$ & $0$& $0$& $1/2$  
&  $1/2$ & $N=1$\\\hline    
$ H$ & $\frac{1}{\beta_1 \cdot \beta_2}(1,\ 2,\ 2,\ 1;\ 1, \ 1)$ & $(bc),bc*$ & $0$   
&  $0$ & $0$ & $0$&  $0$ & $0$ & $N, N^{\prime}$\\\hline\hline
${\ti q}_{R}^1$ & $3({\bar 3},\ 1,\ 1,\ 1; \ 1,\ 1)$ & $ (ah^3)$ & $-1$& $0$ &  $1$ & $0$& $-1/6$ & 
$-2/3$ & $N^{\prime} =1$\\\hline
${\ti q}_{R}^2$ & $3({\bar 3}, \ 1, \ 1, \ 1; \ 1,\ 1)$ & $ (ah^{3*})$ & $-1$ & $0$&  $-1$ &$0$ & $-1/6$ & $-1/3$  & $N^{\prime}=1$\\\hline
${\ti q}_{L}^3$ & $3(3, \ 1,\ 1,\ 1;\ 1,\ 1)$ & $ (ah^4)$ & $1$ & $0$ &$0$ &$-1$ & $1/6$ & $2/3$ & $N^{\prime}=1$\\\hline
${\ti q}_{L}^4$ & $3(3,\ 1, \ 1,\ 1;\ 1,\ 1)$ & $ (ah^{4*})$ & $1$  & $0$ & $0$ & $1$ & $1/6$ & $1/3$ & $N^{\prime}=1$\\\hline
${\ti \phi}_{R}^1$ & $3(1,\ 1,\ 1,\ 1;\ 1,\ 1)$ & $ (dh^3)$ & $0$ & $-1$ & $1$& $0$& $ -1/2$ & $0$ & $N^{\prime}=1$\\\hline
${\ti \phi}_{R}^2$ & $3(1,\ 1,\ 1,\ 1; \ 1,\ 1)$ & $ (dh^{3*})$ & $ 0$& $-1$& $-1$ &$0$ & $-1/2$ & $1$ & $N^{\prime}=1$\\\hline
${\ti \phi}_{L}^3$ & $3(1,\ 1,\ 1,\ 1; \ 1,\ 1)$ & $ (dh^4)$ & $0$ & $1$ &$0$ & $-1$ & $1/2$ & $0$ &  $N^{\prime}=1$\\\hline
${\ti \phi}_{L}^4$ & $3(1,\ 1,\ 1,\ 1; \ 1,\ 1)$ & $ (dh^{4*})$ & $0$ & $1$ & $0$ & $ 1$ & $1/2$ & $-1$ & $N^{\prime}=1$\\\hline\hline
\end{tabular}
\end{center}
\caption{\small Chiral spectrum of the four stack 
D6-brane N=1 Supersymmetric Left-Right Symmetric Standard Model. 
The bottom part 
includes the massive messenger fields that communicate supersymmetry breaking
to the visible N=1 sector. 
Note that the notation L,R at the bottom of the table is  
related to chirality. The gauge group is $U(3)_a \t Sp(2)_b \t Sp(2)_c \t U(1)_{N_{h_3}} \t 
U(1)_{N_{h_4}}$. 
\label{taba1}}
\end{table}

Note that the spectrum of table (\ref{taba1}) is also valid for non-zero NS B-field, that makes the 
tori tilted along the second and the third tori\footnote{As it is already included in the solutions
to the RR tadpoles.}. We also note that the 
higgsino N=1 multiplet states ${\tilde H}_u$, 
${\tilde H}_d$ are part of the 
non-chiral spectrum in the $bc$, $bc*$ sectors respectively and they belong to
of a (2,2) representation of $SU(2)_b \times SU(2)_c$. 
The superpotential part describing the generation of masses in the 
``observable'' part reads
\beq
W^{obs} \sim \l_1 \ Q_L H (i \tau_2) Q_R \ + \ l_{2} \ L H (i \tau_2) R  
\eeq
where $l_1$, $l_2$ may be determined by string amplitudes \cite{ampli}; 
$i \tau_2$ is used for conventional reasons in relation to the 
SM physics (see for example \cite{babu1}).
The representation content of the SM matter is accommodated as 
\beqa
Q &=& 
\left( \ba{c}u\\d \ea \right), \ \ \ R = \left( \ba{c} d^c\\u^c  \ea \right) 
\eeqa
where the Higgs fields are accommodated as 
\beqa
H = \left( \ba{cc} \phi^0_{11} & \phi^{+}_{12} \\
\phi^{-}_{21} & \phi^0_{22} \ea \right)
\eeqa
and may obtain vevs as
$\langle \phi^0_{11} \rangle = u_1 $, $\langle \phi^0_{22} \rangle = u_2 $.

\begin{itemize}

\item  {\large Masses for non-chiral matter  (NCM)}

NCM for the present L-R models appears in 
bifundamentals
arises in $ad$, $ah_1$, $ah_2$, $ah_5$, $dh_1$, $dh_2$, $dh_5$, 
$h_1 h_2$, $h_1 h_5$, $h_2 h_5$ sectors
and it is getting massive by the introduction
of SS deformations as described in the previous section. 
The left-right symmetric models have - compared to the Pati-Salam models
of tables (\ref{appa1}), (\ref{newska}), (\ref{ska})
additional non-chiral sectors coming
from the  $ad$, $dh_1$, $dh_2$, $dh_5$ sectors. The latter extra
matter is also getting massive by the introduced SS deformations in all
tori as there are odd n-wrappings in at least one complex tori in each intersection.

Other NCM that is also present are the one accommodated in the states of table 
(\ref{skla12}) which are getting massive
from the same superpotential terms as the ones of eqn. (\ref{stat1})

\end{itemize}

$\bullet$ {\em \large{Masses for the exotic vector-like multiplets in 
SUSY breaking messenger sector}}
\newline\newline
The present L-R models have also (as the PS models of the previous section) a stable proton as baryon number is a gauged symmetry.
The most obvious phenomenological problem that any string model from D-branes is facing
is the presence of extra massless exotics that survive massless to low energies.

In the present L-R constructions there are superpotential mass terms for all the 
chiral beyond the MSSM fermions -- namely the extra messenger exotics of the bottom
of table (\ref{taba1}) 
${\ti q}_R^1$, ${\ti q}_R^2$, ${\ti q}_L^3$, ${\ti q}_L^4$, ${\ti \phi}_R^1$, ${\ti \phi}_R^2$, 
${\ti \phi}_L^3$, ${\ti \phi}_R^4$. They appear as follows
\beqa
W_{messenger} \ = \ \la_1 \ {\ti q}_R^1 \ {\ti q}_L^3 \  h_{34}^{(2)}   +  \la_2 \ {\ti q}_R^2 \ {\ti q}_L^4 \   h_{34}^{(1)}  
+ \la_3 \ {\ti \phi}_R^1 \  {\ti \phi}_L^3 \  h_{34}^{(2)}   +  \la_4 \ {\ti \phi}_R^2 \  {\ti \phi}_R^4 \  
 h_{34}^{(1)}   
\label{coup1}
\eeqa
The multiplets $h_{34}^{(1)}$, $ h_{34}^{(2)}$ have a gauge singlet scalar direction.
Assuming that it gets a vev 
the couplings (\ref{coup1}) helps the fermion pairs 
${\ti q}_R^1 \ {\ti q}_L^3 $, ${\ti q}_R^2 \ {\ti q}_L^4 $, 
${\ti \phi}_R^1 \  {\ti \phi}_L^3$, ${\ti \phi}_R^2 \  {\ti \phi}_R^4$  
to form massive Dirac mass eigenstates with masses respectively
\footnote{We assume that our vacuum is stable locally} 
\beqa
m_1 =  \la_1 \ \langle 0|  h_{34}^{(2)} |0\rangle \   , \
m_2 =  \la_2 \ \langle 0| h_{34}^{(1)}  |0\rangle \   , \
m_3 =  \la_3 \ \langle 0| h_{34}^{(2)} |0\rangle \   , \
m_4 = \la_4 \ \langle 0| h_{34}^{(1)}   |0\rangle \   .
\eeqa   

We note that the sector made of the H-multiplet preserves a N=2 SUSY. In particular 
it preserves the N=1 SUSY preserved by the observable MSSM and also the $N^{\prime }=1$ 
respected by the messenger sector.

\begin{itemize}
\item {\bf Breaking to the MSSM}

\end{itemize}

In the L-R symmetric models of table (\ref{taba1}), the gauge group $SU(2)_R$ arises
from the c-brane which is placed at a point of the moduli space where it is its own
orientifold image.
The breaking to the $U(2)_R \rightarrow U(1)_c$ and thus the L-R symmetric models 
to the MSSM  can be achieved 
by considering general positions and Wilson lines for brane c \footnote{
See also related comments \cite{lo1}.}
that correspond to geometrical separation along the first torus (where the
branes are parallel) and Wilson line "phases" along the one cycle in the first torus respectively; in turn they are associated to the real and imaginary part of 
the $\mu$-parameter. 
It is is easily 
confirmed that in this case, the (2, 2) N=2 chiral mutiplet $Q_R$ gives rise to 
two N=1 multiplets (2, 1), (2, -1) charged under $SU(2)\t U(1)_Y$ that gets identified with
$U_R$, $D_R$ while the $R$ multiplet in table (\ref{taba1}) gives rise to
$E_R$, $N_R$ as they appear in table (\ref{newtab1}). The extra messenger chiral sector 
remains the one given in the bottom of table (\ref{taba1});
the corresponding states are getting massive once 
the $h_{34}^{(l)}$, l=1, 2 multiplets receive a vev.\newline
Thus at this stage -- being at the string scale -- only the MSSM multiplets, seen at the top part of table (\ref{newtab1}), survive massless. The only question
remaining at this point is to show in which way the MSSM spartners receive masses after 
susy breaking. This may be achieved in the next section by the use of D-term breaking.


\subsection{{D-term Supersymmetry breaking and Gauge Mediation}}

In the usual N=1 SUSY models of gauge mediation \cite{gm}, the spectrum consists
of the MSSM together with an extra hidden sector $G_H$ that contains vector-like 
messenger particles $\phi_H$, ${\bar \phi}_H$ charged
under the observable MSSM and the $G_H$ gauge group. 
SUSY is broken in the hidden sector when a spurion superfield $X_H$ gets a vev 
through the superpotential 
\beq
W^{(N=1)} \ = \ W_{obs} \  + \ \phi_H  {\bar \phi}_H  X_H 
\eeq
and where the $W_{obs}$ and the susy breaking part of the superpotential W respect the same N=1 susy. In field theory terms, the squarks and sleptons get masses from two loop
effects while gauginos get masses from one loop effects where 
messenger states circulate in the loops.
On the other hand in the context of string theory models from intersecting branes 
it was argued in a 
field theoretical basis that in non-susy models - localized within 4D toroidal 
orientifolds of IIA with intersecting D6-branes - that exbibit the 
quasi-susy spectrum (By definition models where MSSM 
particles are subject to different N=1 supersymmetries) similar effects are 
expected \cite{Iba}.

The supersymmetry observed in the gauge mediated construction of 
table (\ref{taba1}) can be broken by slightly varying the complex 
structure of the models. In the low energy effective theory this procedure can be seen
as turning on Fayet-Iliopoulos terms for the U(1) fields \cite{Iba}.
Such a procedure has been applied to quasi-susy models in \cite{quasi}.
In the present modes, which can avoid the non-chiral fermions once Scherk-Schwark 
breaking is included it appears that SS deformations 
do not affect the angles between the branes in the RR tadpoles. In general 
for two D6-branes, 
that exhibit N=1 SUSY limit at their intersction,
the scalar potential contribution appears as 
\beq
V_{FI} = \frac{1}{2g_a^2}( \sum_i q_a^i |\phi|^2 + \xi_a )^2 \ .
\eeq
Lets us be more specific and consider a slight departure in the complex strusture of eqn.(\ref{susy}) 
that defines the supersymmetric limit   
\beq
\beta_1 {\ti \chi}_2^{\prime} = \beta_1 \chi_2 + k_a \  ,
\eeq
where $k_a << 1$ and where $k_a$ could be also written as 
$k_a = \beta_1 p_a$, where $p_a$ represents the actual variation in the 
complex structure. When the torus is untilted then $k_a \equiv p_a$.
Lets us calculate as an example the angle \footnote{We use the D6-brane 
wrappings as they are seen in table (\ref{skaB})  of appendix B. } 
between the D6-branes $a$, $b$ along the second 
torus (seen in table (\ref{talo1}). 
It is given by
\beqa
\theta_{ab}^{2} & = & \ tan^{-1} \left( (3 \epsilon \rho^2) \   \b_1 \chi_2^{\prime} \right) \approx
tan^{-1} \left( (3 \epsilon \rho^2) \   \b_1 \chi_2 \right) \  +  \ \frac{(3 \epsilon \rho^2) k_a}
{1+ (3 \epsilon \rho^2 \b_1 \chi_2 )^2} \\
& = & a_1  \ +  \ \delta_a \ ,
\label{angle1}
\eeqa  
where $a_1 = tan^{-1}(3 \epsilon \rho^2 \b_1 \chi_2)$, $\delta_a = ((3 \epsilon \rho^2) k_a )/(1+ (3 \epsilon \rho^2 \b_1 \chi_2 )^2) $.

\begin{table}
[htb]\footnotesize
\renewcommand{\arraystretch}{1.3}
\begin{center}
\begin{tabular}{||c||c|c|c||} 
\hline
\hline
$\ Brane $ & $\ (\theta_a^1)\ $  & $\ (\theta_a^2) \ $  & $\ (\theta_a^3) \ $\\
\hline\hline
 $a$  & $0$  & $a_1 + \d_a $ & $- a_1$\\
\hline 
$b$  & $\frac{\pi}{2} $ & $0$ & $-\frac{\pi}{2}$ \\
\hline
$c$ & $\frac{\pi}{2} $ & $-\frac{\pi}{2}$  & $0$\\    
\hline
$d$ & $0$  & $a_1 + \d_d$ &   $-a_1$ \\\hline\hline
$h_3$ &  $\frac{\pi}{2} $ & $0$ & $\frac{\pi}{2}$ \\\hline
$h_4 $ &  $\frac{\pi}{2}$ & $\frac{\pi}{2}$ & $0$ \\\hline
\end{tabular}
\end{center}
\caption{\small Angle structure between the D6-branes
and the orientifold axis for the left-right symmetric model. 
We exbibit only branes that give rise to chiral matter structure.
\label{angle}}          
\end{table}

\begin{table}
\renewcommand{\arraystretch}{1.8}
\begin{center}
\begin{tabular}{||c||c|c|c|||||} 
\hline
\hline
$Sector$ & $(\theta_a^1 , \ \ \theta_a^2 , \ \ \theta_a^3 )$ & sparticle & (mass)$^2$\\
\hline\hline
 $(ab)$ & $( -\frac{\pi}{2}  ,\  \a_1 + \d_a , \ - a_1 + \frac{\pi}{2} )$  & $1 \t {\tilde Q}_L$ 
& $\frac{M_s^2}{2}(\d_a)$\\
\hline
$ac$ & $( -\frac{\pi}{2}  ,\  \a_1 + \d_a + \frac{\pi}{2} , \ - \a_1 )$ & $3 \t {\tilde Q}_R$ 
&  $\frac{ M_s^2   }{2 }(\d_a )$     \\\hline
$db$ & $(- \frac{\pi}{2} , \ \a_1 + \d_a , \ -  a_1 + \frac{\pi}{2} )$ & $3 \t {\tilde L}$ 
&    $\frac{  M_s^2 }{2 }(\d_a)$     \\\hline
$dc$ & $(   - \frac{\pi}{2} , \ \a_1 + \d_d +  \frac{\pi}{2} , \ - \a_1   ) $ 
& $3 \t {\tilde R}$  &     $\frac{M_s^2 }{2}(\d_a)$                  \\\hline
$bc $  & $( 0 \ , \  \ \frac{\pi}{2} , \ \  - \frac{\pi}{2}  )$   &   
$\frac{1}{\b_1 \b_2} \t {\tilde H}$  &   $\frac{M_s^2 \ l_2}{4  \pi^2}$    
\\\hline\hline\hline\hline
$ah_3$  &   $( -\frac{\pi}{2}  \ ,  \a_1 + \d_a    ,  \  - \a_1 - \frac{\pi}{2})$ &  
$3 \t {\tilde q}_R^1 $   &     $\frac{M_s^2 }{2 }( \d_a)$   \\\hline
$dh_3$  &  $(  -\frac{\pi}{2}  \ ,  \a_1 + \d_d , \  - \a_1 - \frac{\pi}{2}    )$ 
& $3 \t {\tilde \phi}_R^1 $ &  $\frac{M_s^2 }{2 }(\d_a)$  \\\hline
$ah_3^{\star}$  &  $( \frac{\pi}{2}  , \  \a_1 + \d_a , \   - \a_1 +  \frac{\pi}{2})$ 
& $3 \t {\tilde q}_R^2 $   &   $\frac{M_s^2 }{2 }(\d_a )$    \\\hline
$dh_3^{\star}$  & $(  \frac{\pi}{2}  , \  \a_1 + \d_d , \   - a_1 +  \frac{\pi}{2} )$ 
& $3 \t {\phi}_R^2 $  &   $\frac{M_s^2 }{2 }(\d_a)$     \\\hline
$ah_4$  &  $(  \frac{\pi}{2} , \  \a_1 + \d_a - \frac{\pi}{2}  , \   - a_1 )$ 
& $3 \t {\tilde q}_L^3    $ &     $\frac{M_s^2 }{2 }( \d_a)$   \\\hline
$dh_4$  &  $( \frac{\pi}{2}  , \  \a_1 + \d_d - \frac{\pi}{2} , \   - a_1 )$ & 
$3 \t {\tilde \phi}_L^3 $  &  $\frac{M_s^2 }{2 }(\d_a)$  \\\hline
$ah_4^{\star}$ &  $(  \frac{\pi}{2}  , \  \a_1 + \d_a + \frac{\pi}{2}  , \   - a_1    )$ & $3 \t {\tilde q}_L^4$   &   $\frac{M_s^2 }{2 }(\d_a )$   \\\hline
$dh_4^{\star}$  &  $(   \frac{\pi}{2} , \  \a_1 + \d_d + \frac{\pi}{2}  , \   - a_1     )$ &    $3 \t {\tilde \phi}_L^4$     &  $\frac{M_s^2 }{2 }(\d_a )$   \\\hline
\end{tabular}
\end{center}
\caption{\small Universal Squark and Slepton masses from FI-terms in the Left-Right 
Symmetric model.
\label{talo1}}          
\end{table}

\begin{itemize}
\item {\bf {\large Universal squark and slepton masses}}
\end{itemize}

We observe that all string (mass)$^2$ of the squark and slepton masses are positive 
and hence colour and charge 
breaking minima may be avoided. The mass$^2$ value of the Higgses does not get
corrected by the variation in the complex structure (or the variation of the magnetic fields 
in the T-dual language with D9-branes and fluxes) but retains only its dependence on the 
distance between the parallel D6-branes in the second tori. We notice that between the 
SU(2) b, c branes there is a N=2 preserving sector which possesses two scalar Higgs states and 
also
two fermionic Higgsino partners, all with masses$^2$ $\frac{M_s^2}{4 \pi^2}l_2 $ at tree level.
 The masses of the Higgsinos depend on the distance $l_2$ between the branes b,c 
which is a open string 
modulus and thus can be be made small. 
Furthermore, the sparticle masses get a 
universal value and as a result we expect that flavour changing neutral 
currents (FCNC) to be strongly suppressed \footnote{  
Universal soft terms masses for the sparticles also
appear (as a result of the hypothesis of parametrizing the unknown 
susy breaking effects in the scalar potential expansion) on another occasion 
in string theory, in N=1 heterotic orbifold compactifications of 
the
heterotic string where the dilaton dominant source of susy breaking \cite{dila}; and also in models with fluxes - 1st ref. of \cite{openmoduli}.}.
The squarks and slepton masses seen in table (\ref{talo1}) are the exact 
string expressions as they are calculated by the use of the appropriate 
mass operators (\ref{mainter}). These masses could be calculated in 
the limit of small 
deviation in the complex structure of the second torus. 
In this case, one can also calculate the field theory limit of the 
supergravity (sugra) approximation $k_a << 1 $ (e.g. if for the radii 
$R_1 >> R_2$ ) that is coming from Fayet-Iliopoulos (FI) terms
and find that the string and sugra
approximations of the scalar masses at the intersections coincide \cite{Iba}.
The string expression is more generic. Thus the scalar masses at an intersection
between two branes coming in the limit of small variation in the complex structure {\em could be understood as coming
 from FI terms} in the effective theory.


\section{{Split Supersymmetry in String theory and MSSM}} 

The Split supersymmetry scenario (SSS) \cite{split0} was proposed as an alternative 
possibility for generating a signal for LHC. As in a global susy or supegravity 
context 
there is no underlying principle for constraining the mass of the particles,  
in the original SSS appearance \cite{split0}, 
it was assumed that spartners of the MSSM are massive at the
UV high scale $m_s^H$ where supersymmetry is getting broken, while gauginos and 
Higgsinos are the TeV scale in order to retain the unification of the gauge couplings at the 
$10^{16}$ GeV.
Furthermore it was assumed \cite{split0} that one of the Higgs 
doublets of the MSSM is finely tuned to be light and below $m_s^H$. 
Exactly at the same  
time it was argued that in the context of string theory split susy models
should have slightly different characteristics respectively \cite{split1, split2, orbiko}. 
These properties that can still keep the nice features that are needed for realistic string model building
are : proton stability, partial unification of two of the gauge couplings at the high 
scale \cite{split1, split2, orbiko} and either light gauginos \cite{split1}; and higgsinos that be 
either light \cite{split1}
or anywhere in the range $M_z$ to $M_s$\cite{split2, orbiko}; and in all cases massive squarks and 
sleptons \cite{split1, split2, orbiko}. 
In stringy MSSM models of this work proton is stable as baryon number is a gauged symmetry.  Next we discuss the 
rest of the {\em Stringy Split SUSY} criteria set out in \cite{split1, split2, orbiko} and which concern
gauge coupling unification of SU(3) and SU(2) gauge couplings accompanied by 
appropriate Weinberg angle sin$^2 (\theta_W)$ at the unification scale; it turns out that gaugino and higgsino 
masses are 
different from the standard split sysy scenario of the local theory \cite{split0}.

\subsection{\bf Gauge Coupling Unification}

Nesessary ingredient of any split theory is that the successful prediction of unification
of gauge interactions \cite{split0} is retained.   
In \cite{split}, based on examples on D-brane inspired models, 
it was argued that 
models of split susy in a string theory content should also accommodate the successful GUT 
prediction $sin^2(\theta_W)= 3/8$ with equal SU(3) and SU(2) gauge couplings
at the unification scale. Explicit realizations of 
non-supersymmetric string models with intersecting D6-branes
where $sin^2(\theta_W)= 3/8$ and equality of SU(3) and SU(2) gauge couplings 
had appear in \cite{split2}, \cite{orbiko}. The latter models are based on $Z_3 \t Z_3$ orientifold 
compactifications of type IIA theory.
These models are non-susy and have no
supersymmetry preserved at the different intersections in sharp contrast
with the models presented in this work where the appearance of N=1 susy on 
intersections is explicit.
In the present models before the breaking of the left-right symmetric models to the 
MSSM at $M_s$,
the gauge group at the observable sector is an $SU(3)_c \t Sp(2)_W \times U(1)_a \t U(1)d \t SU(2)_R$.
When all the tori along the three complex dimensions are orthogonal 
the unification of the gauge couplings of the ``observable'' SM spectrum 
is described by the wrappings numbers of the top part of table (\ref{newska}) when $\beta_1 = \beta_2 =1$, and    
has been studied in \cite{blu}. In this case it was found that 
$sin^2 (\theta_W) = \frac{3}{5}$ at the string scale as in SUSY SU(5). \newline
In this section, we study the gauge coupling unification of the MSSM's 
coming from the breaking of  
$SU(4)_C \t SU(2)_W \t SU(2)_R$; the tori could be also tilted. 
The general background is a four dimensional IIA orientifold compactification on a six dimensional tori in the 
presence of D$6_a$-brane wrapping at angles on a general factorizable 3-cycle 
described by the wrappings $(n, m)_a$.
Using the general RR solutions of table (\ref{ska}) we find 
that the volumes obey \cite{cre}
\beq
V_a \ = \ V_d \ ,\ V_b \ = \ V_c
\label{vol}
\eeq
The relation of gauge couplings in terms of wrappings and the complex 
moduli $\chi_i$ along the three tori, 
are given by the generic relation \cite{blu}
\beq 
\frac{1}{a_a} = \frac{1}{k_a}\cdot \frac{M_{pl}}{2 \sqrt{2}} \cdot \frac{V_a}{ M_s  \sqrt{V_6}},
\eeq
where 
\beqa
a_a^{-1} =  \frac{1}{\sqrt{2}k_a }\frac{M_{M_{pl}}}{M_s} \times &  \nonumber\\
 \left( (n^1 n^2 n^3)_a \sqrt{\frac{1}{U_1 U_2 U_3}} - (n^1 m^2 m^3 )_a \sqrt{\frac{U_2 U_3}{U_1}} -
(m^1 n^2 m^3)_a \sqrt{\frac{U_1 U_3}{U_2}} -
(m^1 m^2 n^3 )_a \sqrt{\frac{U_1 U_3}{U_2}}  \right) && \nonumber\\
\label{ga1}
\eeqa
and the value of $k_a$ is set out by the gauge group on the brane $a$ and is $k_a =1$
when the gauge group is a U(N) one and  $k_a =2$ when the gauge group is an Sp(N).
we derive in the general case of tilted tori, the gauge couplings at the string 
scale of the MSSM.  We use the more general solution to the RR tadpoles seen in 
table (\ref{ska}) of appendix B for the 
Pati-Salam models that accommodates the parameter $\rho=1,1/3$. 
We set 
\beq
\frac{1}{\sqrt{2}k_a }\frac{M_{M_{pl}} }{M_s} = K_0 \ ,\ \  \b_2 \chi_2 = \b_3 \chi_3  
\eeq
The gauge couplings are 
\beqa
\frac{1}{a_s}&=& K_0 \frac{\sqrt{\beta_1}}{\sqrt{\beta_2}} \frac{1}{\sqrt{\chi_1}} \left(  
 \frac{1}{\rho^2}\frac{1}{\chi_3} \ + \ 9 \rho^2 \beta_2 \chi_3  \right)
\nonumber\\
\frac{1}{a_W}&=& \frac{K_0}{2}\frac{\epsilon}{\sqrt{ \beta_1 \beta_2}}\sqrt{\chi_1} \nonumber\\
\frac{1}{a_c}&=&K_0 \frac{\epsilon}{\sqrt{\beta_1 \beta_2}}\sqrt{\chi_1} 
\eeqa
where $a_c = \frac{1}{2}a_W$ and the gauge group for general positions and Wilson 
lines for c-brane is an $SU(3)_C \t Sp(2)_W \t U(1)_Y$ ($Sp(2) \equiv SU(2)$).

The gauge couplings of the $SU(3)_c \t SU(2)_w \t U(1)_Y$ gauge group are
\beq
a_s = a_d \ , \  \ a_c = \frac{1}{ 2 } a_b 
=  \frac{1}{2 } a_w
\eeq
Also since
\beqa
\frac{1}{a_Y}  &=&  \frac{1}{6}\frac{1}{a_s} \ + \ \frac{1}{2}\frac{1}{a_c} \  
+ \ \frac{1}{2}\frac{1}{a_d}
\eeqa 
we find that at the string scale the gauge couplings obey
\beqa
\frac{1}{a_Y} &\stackrel{M_s}{=}& \frac{2}{3} \frac{1}{a_s} \ + \ \frac{1}{a_w} \ .
\label{asa1} 
\eeqa
From (\ref{asa1}) we derive the value of the Weinberg angle at the string scale 
\beq
sin^2(\theta_W)_{M_s} = \frac{3 a_s}{6 a_s + 2 a_w} \ .
\label{sin}
\eeq
The free parameters in this equation are the values of the complex structure 
moduli $\chi_i$, i=1, 2, 3. 
Complex structure moduli, as was first noticed in \cite{kokos1} [see eqn.(4.37) in hep-th/0203187] in the context of 4D toroidal orientifold intersecting brane worlds \cite{lustr}, 
the supersymmetry conditions in chiral fermion intersections could be all fixed 
to a certain value without the presence of fluxes.
These models were based on 
Pati-Salam $SU(4) \t SU(2)_L \t SU(2)_R$ type of constructions and in the same
4D string backgrounds of the present work.

\begin{itemize}

\item {\large{Fixing complex structure moduli via $sin^2(\theta_w)$ at the String scale $M_s$}} 

\end{itemize}

In the present constructions complex structure moduli could be further 
constrained - but not completely fixed - using the values of the gauge couplings at the string 
unification scale. 
From (\ref{sin}) we observe that successful unification relations such that of SU(5) GUT could be also 
derived within the framework of Pati-Salam GUTs from intersecting branes.  
Hence by demanding that 
\beq
a_s \ = \ a_w \ ,
\label{labe}
\eeq
using appendix C, we can reproduce the GUT 
predictions for sin$^2(\theta_w)$ at the unification/string scale as
\beq
a_a = a_w  = \frac{5}{3} a_Y \ , \  \ sin^2(\theta_w)(M_s) = \frac{3}{8}
\label{su5}
\eeq
So the MSSM models behave as a hidden SU(5) at the unification GUT/String scale of the SU(3) 
and SU(2) couplings. \newline
Eqn. (\ref{labe}) could be used to constrain the complex structure moduli $\chi_1$ - along the first torus - 
in terms of the values of the rest of moduli along the 2nd and 3rd tori that are related
by N=1 SUSY as in (\ref{susy}).
From (\ref{labe}) we get a second order equation in the form
$\alpha \chi_1^2 + b \chi_1 + \gamma = 0$, on the $\chi_3$ modulus,
where
\beq
{\ti a} = 9 \rho^4 \beta_2 \ , \ {\ti b} = \chi_1 \rho^2 (2 \beta_1 \sqrt{\beta_2} )^{1/2} , \ \ 
 {\ti \gamma} = 1
\eeq
Demanding positivity ${\ti b}^2 - 4 {\ti a}{\ti \gamma}$ of the 
square root (to demand only real values for the 
modulus $\chi_1$) so at least one root is positive \footnote{namely the 
$-{\ti b}/2{\ti a} + 1/(2{\ti a})\sqrt{{\ti b}^2 - 4 A {\ti \gamma}} $}, we get the constraint
\beq
\chi_1 = \frac{R_2}{R_1} > \frac{ \beta_1 \beta_2 }{3} 
\eeq
The present models reveal the presence of the successful SU(5) GUT prediction $sin^2(\theta_W) =3/8$,  when all tori were orthogonal 
(this was also noticed in \cite{blu}
\footnote{See also first ref. of \cite{bluo1} for an attempt to construct N=1 Pati-Salam MSSM-like models 
where however sin$^2 (\theta_w)$ at the GUT/string scale does not possess a hidden SU(5), e.g. $sin^2 (\theta_w ) \neq 3/8$.}) and also when
tilted tori are involved. It appears that the presence of a "hidden" SU(5)
inside Pati-Salam models that is behind
(\ref{su5}) is independent of the existence or not of tilted tori.

\begin{itemize}

\item {\large{Gauge coupling unification at $M_s$}} 

\end{itemize}
At this part, we determine the value of the string scale in which the gauge couplings of the
SM unify \footnote{We briefly reproduce a relevant discussion from \cite{blu}.}.
The tree level relations of the gauge couplings of the SM at the string scale read
\beqa
\frac{1}{a_s(\mu)}=\frac{1}{a_s} + \frac{b_3}{2 \pi}ln(\frac{\mu}{M_s})\nonumber\\
\frac{sin^2 (\theta_w)(\mu)}{a(\mu)}=\frac{1}{a_w} + \frac{b_2}{2 \pi}ln(\frac{\mu}{M_s})\nonumber\\
\frac{cos^2(\theta_w)(\mu)}{a(\mu)}=\frac{1}{a_Y} + \frac{b_1}{2 \pi}ln(\frac{\mu}{M_s})
\eeqa
By using the relation (\ref{susy})we obtain that at the string scale 
\beq
\frac{2}{3}\frac{1}{ a_s(\mu)}  \ + \  \frac{2 sin^2(\theta_w(\mu)-1}{2\pi}=
\frac{ {\hat B}  }{2\pi}\ln\left( \frac{\mu}{2\pi}  \right)
\label{refs1}
\eeq
where 
\beq
{\hat B} = \frac{2}{3}b_3 + b_2 - b_1
\eeq
After the breaking of the left-right symmetric model at the MSSM, the massless content
of our models at the string scale is the one given by the MSSM in addition to the 
extra matter as it seen in the bottom of table (\ref{taba1}).
As the extra exotic non-chiral matter is getting massive by the vevs of the flat directions 
$h_{34}^{(l)}$, l=1,2
which are of the order of the string scale, at $M_s$ only the MSSM remain massless for which
${\hat B} =12$.
Thus the unification scale (US) - at which the SU(3) and SU(2) couplings unify for the 
present MSSM models is just 
\beq
M_s = M_{GUT} = 2.04\t 10^{16} \ GeV
\eeq 
{\em This scale is independent of the number of Higgsino multiplets in the MSSM as their 
contribution within the combination ${\hat B}$ cancels out } as
\beq
b_3 = - 2n_G + 9 \ , \  b_2 = -2 n_G - 6 - n_u \frac{1}{2} - n_d \frac{1}{2} \ , \ 
b_1 = - n_G\frac{10}{3} - n_u \frac{1}{2} - n_d \frac{1}{2}
\eeq
and $n_G = 3$ in the present models.

\subsection{\bf Gaugino, Higgsino and Dark Matter Split SUSY candidates}

\begin{itemize}
\item {\bf \large{Gaugino masses}}
\end{itemize}
In \cite{split1} it was argued - following arguments on Scherk-Schwarz compactifications
in the absence of branes intersecting at angles \cite{SSta1}
\footnote{where repeating the argument \cite{SSta1} $m_{1/2} \sim m^3_{3/2} / 
M_p^2$, $m_{1/2} 
\sim$ TeV when  $m_{3/2} \sim 10^{13-14}$ GeV. Notice that it is also allowed
$m_{1/2} 
\sim 10^{10}, 10^{19}$ GeV when $m_{3/2} \sim 10^{16},  10^{19}$ GeV respectively.
 } -  
that gauginos in string theory should receive TeV scale masses.   
On the contrary in \cite{split1, orbiko}, it was suggested 
that gauginos should only receive masses of the order of the string scale even 
though at tree level in intersecting brane models appear to be massless. 
The latter suggestion originated from field theory arguments for intersecting 
brane models in toroidal orientifolds \footnote{As they appear in the appendix 
of \cite{imr}.}, 
where the $D6_a$ gauginos 
receive non-zero masses from one loop corrections 
from massive N=1 
hypermultiplets running 
in the loops, while the extension 
to the $Z_N, Z_N \t Z_M$ orbifolds case comes by considering the contribution 
from the different intersections
when taking into account the orbits in the orbifold brane structure. 
The stringy one loop correction for the toroidal orientifold case was verified 
recently by a string calculation \cite{anto6}.

\begin{itemize}
\item {\bf \large{Higgsino masses}}
\end{itemize}
It was suggested \cite{split1} that in the context of string theory,
a necessary condition for having a light Higgs multiplet in the spectrum 
would be to have
unbroken supersymmetry in the Higgs sector even after supersymmetry breaking, 
that is tree level massless Higgs multiplets
in the spectrum. Moreover it was also suggested \cite{split1} that it was likely
- in the sense when 
such explicit models will be 
constructed  -  given the similarity of the Higgs doublets to sleptons that even sleptons could
be found massless at tree level.
Different suggestions for the exact mass of the Higgsinos in string models had appear in \cite{split2, orbiko}.
In \cite{split2, orbiko} it was suggested  
that in non-susy models with no susy at intersections on toroidal orientifolds Higgsinos could be in principle {\em anywhere} 
between the weak and the string scale. 
In the latter orbifolded orientifolds the chiral spectrum consists of only 
\footnote{neglecting the non-chiral matter}
the MSSM fermions - in addition to three pairs of massive 
non-chiral colour exotics that become massive from existing Yukawa couplings  
- and where the Higgsinos make Dirac pairs and couple to gauge singlets at tree level. 
See for example section (8.1) in \cite{split2} 
and eqn. (6.7) in \cite{orbiko}.
Thus the models of \cite{split2, orbiko} - where supersymmetry is already broken by construction at the
string  scale $M_s$ in the open string sector where the SM is localized, accommodate 
all the split susy criteria namely: \newline
a) equality of SU(3) and SU(2) gauge couplings and correct
$sin^2 \theta_W = 3/8$ at the unification/string scale, \newline 
b)accommodating 
massive gauginos and higgsinos that could be 
anywhere from $M_z$ to $M_s$ ($M_z \leq \mu \leq M_s$).  \newline
Thus interesting conclusions on the magnitude of Higgsino masses in string models
with the split susy characteristics were obtained. 
However since in these models susy is broken by construction 
in the open sector high at $M_s$ and there is no explicit N=1 SUSY preserved in 
their Higgs sector, 
the solution of the gauge hierarchy problem in the Higgs sector
of these models is not guaranteed \footnote{In addition the models of 
\cite{split2, orbiko} have also present 
the non-chiral matter that appears in models with intersecting D6-branes.}. \newline
On the contrary in the MSSM models of this work - where N=2 supersymmetry is preserved 
in the N=2 sector - after
the breaking of supersymmetry, Higgsinos do not receive any 
correction (modulo the 
distance $l_2$ between the branes b, c that can be made very small and is an open 
string modulus) at tree 
level due to the D-term FI breaking as
we have seen in the previous section.
They do however receive a one loop correction
of the order $\sim \delta_a^2 M_s $ \cite{anto6}.

\begin{itemize} 

\item {\bf \large{Dark matter candidates}}

\end{itemize}

One of the most difficult cosmological problems is the composition of 
the present day dark matter (DM) density $\Omega_{DM}$ in the universe. 
Weakly interacting 
massive particles (WIMPS) are good candidates for generating $\Omega_{DM}$. 
In the context of supersymmetric or SUGRA theories where the breaking of 
supersymmetry used to be at 
the TeV to avoid the hierarchy problem, neutralinos $X^{neu}$ 
\footnote{we denote them by capital $X^{neu}$ to avoid confusion with $\chi$ which is used
to describe the complex structure.}
are popular candidates for generating the cold dark matter of the 
universe (see \cite{popu}). They 
are expected to be non-relativistic in the present epoch and are eigenstates of a mixture of
 binos ${\ti B}$, winos ${\ti W}_3$ (the superpartners of B, $W_3$) and neutral 
components of higgsinos where
\beq
X^{neu} = N_{11}{\ti B} +  N_{11}W_3 +  N_{13} H_1^0 +  N_{14} H_2^0
\eeq
E.g. the higgsino fraction is defined as $N_{H} = | N_{13}|^2 + | N_{14}|^2$  and is used to 
distinguish the amount of higgsinos mixing in the $X^{neu}$ composition, e.g.
if one could find that e.g. $N_{H} > 0.9$ then such a mixing could be considered as mostly higgsino. 
\newline
In the present models, the introduction of SS breaking which makes massive all the non-chiral matter fields of 
the models, make also massive and of the order of the string scale the gauginos which belong to branes with odd "electric" n-wrappings.
that are charged under the SM gauge group.  Which gauginos could 
become massive also depends on 
the number of tilted tori (or the NS B-field) that enter the RR tadpoles.
Gauginos that are associated with a D6-brane for which SS 
deformations act in some complex tori receive 
\footnote{Non-chiral matter also receives a similar mass.} a string scale mass of order
\beq 
m \sim \frac{1}{2\sqrt{n_i^2 R_1^2 + m^2 R_2^2}}
\label{contribu1}
\eeq
where $(n,m)$ the wrapping numbers of the tori for which the SS deformation
is acting. 

$\bullet$ {\em Neutralinos} : Only Higgino $\beta_1 =  \beta_2 = 1$

In the PS models of section 3 and the left right symmetric models of section 4, all gauginos may receive masses from 
SS breaking due to their odd-wrappings. Thus the neutralino could be only 
made of Higgsinos.  Also charginos are made only from only charged Higgsinos.

$\bullet$ {\em Neutralinos: only Higgsino-Wino}; $\beta_1 = 1/2$; $\beta_2 =1$

In this case, the second torus is tilted and the b-gauginos are massless as they receive no 
mass from SS deformations while we have present two pairs of tree level massless Higgsinos 
$(H_u, H_d)$.
Thus at low energy neutralinos are made 
only from neutral components of higgsinos 
and "electroweak" Winos. 
Mixing of the higgsinos with the Winos perhaps is necessary in order to 
break the degeneracy of the two Dirac type lightest Higgsino newtralinos and to scatter 
them inelastically off the nucleus \cite{smith}. 
Also {\em charginos} which in general is a mixture of the 
charged components of higgsinos and charged gauginos could be also made  
from charged higgsinos and Winos.
Binos may be massive as their gauginos are associated with the ones coming from 
the adjoint breaking of the the U(4) Pati-Salam branes that create the left-right
symmetric models.

The introduction of the tilt in the second torus  
reintroduces the massless non-chiral fermions, as the SS deformation no more
affects their masses. Thus massless adjoint fermions from the weak brane are reintroduced 
and at this level we have no suggestion how to give them a mass.
However, the non-chiral non-adjoint matter 
from the intersections
$bh_1$, $bh_3$, $bh_5$, $ch_1$, $ch_4$, $bh_5$; where we denote the associated fermions as  
$\chi^{( bh_1 )}$, $\chi^{( bh_3 )}$, $\chi^{( bh_5 )}$, $\chi^{( ch_1 )}$, $\chi^{(ch_4 )}$, 
$\chi^{(bh_5 )}$
are getting massive from the superpotential couplings \footnote{Superscripts
denote the corresponding intersection.}
\beq
\chi^{ (bh_1 )} \chi^{( bh_3 )} \chi^{( h_1 h_3 )} ; \
\chi^{ (bh_5 )} \chi^{( ch_4 )} \chi^{( bc )} \chi^{( h_4 h_5 )} ; \
\chi^{ (ch_4 )}_{(0;0,1)} \ \chi^{( ch_5 )}_{(0;0,1)} \ \chi^{( h_4 h_5^{\star} )}_{(0;-1,-1)} \ , \
\label{laster1}
\eeq
where $\chi^{( h_1 h_3 )}$, $\chi^{( h_4 h_5 )}$,  $\chi^{( h_4 h_5^{\star} )}$, the N=2 matter that possess
gauge singlet directions under the SM hypercharge. In the last term in 
(\ref{laster1}), we have denote by underscript the corresponding charges under the
gauge groups of the $(c;h_4, h_5)$ D6-branes respectively. By $\chi^{( bc )}$ we denote the difundamental Higgs multiplets  
appearing as a basic ingredient in the spectrum of any PS or left-right symmetric model.We also mention of the possibility for the gravitino to be 
the LSP as it happens in local theories of gauge mediation with low scale 
susy breaking. See also \cite{split} for some relevant work in a split 
susy non-string context. Further consequences of the present models
for dark matter will be considered elsewhere.


\section{Addition of RR, NS and Metric fluxes}

In this section, we will examine the classes of Pati-Salam models of
this work, by studing
their low energy effective action in the presence of background fluxes. In particular
we study the question of moduli stabilization in the presence
of RR/NS and also metric fluxes as they have been recently 
incorporated in \cite{vz} and further studied in \cite{cfi}.      
In type IIA Calabi-Yau orientifolds such as the $T^6$ torus the introduction
of fluxes makes possible to fix most of the moduli of the theory.  
Given the unique string Pati-Salam models of the previous sections that leave only the MSSM at the
string scale the question is if these models can be also reproduced in the context of fluxes where 
also moduli could be fixed.

\subsection{Preliminaries - RR tadpoles}

The general superpotential 
involving RR/NS and metric fluxes is the sum of the terms generating 
the K\"ahler and the complex structure moduli, $W_K$, $W_c$ respectively.
We define
\beq
W_K \ =  \  \int_Y e^{J_c} \wedge {\bar F}_{RR}
\eeq 
where ${\bar F}_{RR}$ represents a sum of even RR fluxes. 
The complex structure moduli $W_c$ is a sum of two terms, the first one 
that \footnote{
\beqa 
\wt{T}_I & = &(i, T_1, T_2, T_3)
\quad ; \quad A_{IJ} = \bmat{cccc} \!\!\! -h_0 & h_1 & h_2 & h_3 \\ \,
a_1 & b_{11} & b_{12} & b_{13} \\ \, a_2 & b_{21} & b_{22} & b_{23} \\
\, a_3 & b_{31} & b_{32} & b_{33} \emat \label{amatrix} \\[-0.4cm]
\wt{U}_I & = &(S, U_1, U_2, U_3) \ .  \nonumber 
\eeqa 
} 
corresponds to the one generating a flux due to the NS $\ov{H}_3$ and the second one that 
is generated 
by the metric fluxes inside $dJ_c$;
  $W_c$ was computed \footnote{Detail definitions can be found e.g. in \cite{cfi}.} in \cite{vz} as
\beq
W_c = \int_Y \Omega_c \wedge (\ov{H}_3 + dJ_c) =  - \sum_{I,J = 0}^3 A_{IJ} \wt{T}_I \wt{U}_J
\eeq

In the case of our models that are accommodated in the presence of RR/NS and metric fluxes -- within 
four dimensional toroidal compactifications of type IIA theory -- 
(and intersecting D6-branes) the full superpotential is computed to be \cite{cfi}
\beqa 
W & = & W_c + W_ Q = e_0
+ ih_0 S + \sum_{i=1}^3 [(ie_i - a_i S - b_{ii}U_i -\sum_{j\not= i}
b_{ij}U_j)T_i - i h_iU_i] \nonumber \\[0.2cm] & - & q_1 T_2 T_3 -q_2
T_1 T_3 -q_3 T_1 T_2 + i m T_1 T_2 T_3 \ .
\label{wa}
\eeqa 
This is a result also obtained in \cite{vz}.
Recall that the metric fluxes are constrained by the Jacobi identities which imply 
the twelve constraints 
\beqa 
b_{ij}
a_j + b_{jj} a_i & = & 0 \quad ; \quad i \not= j \nonumber \\[0.2cm]
b_{ik} b_{kj} + b_{kk} b_{ij} & = & 0 \quad ; \quad i \not= j \not= k
\ .
\label{jacb} 
\eeqa 
The RR tadpole conditions become 
\beqa 
\sum_a  N_a n_a^1 n_a^2 n_a^3 + \frac12(h_0 m + \sum_{i=1}^3 a_i q_i ) & = & 16 \  ,
\nonumber \\
\sum_a  N_a n_a^1 m_a^2 m_a^3 + \frac12(h_1 m - \sum_{i=1}^3 q_i b_{i1} ) & = & 0  \  ,
\nonumber \\
\sum_a  N_a m_a^1 n_a^2 m_a^3 + \frac12(h_2 m - \sum_{i=1}^3 q_i b_{i2} ) & = & 0  \  , \nonumber\\
\sum_a  N_a m_a^1 m_a^2 n_a^3 + \frac12(h_3 m - \sum_{i=1}^3 q_i b_{i3} ) & = & 0 \ . 
\label{lio1}
\eeqa 
Hence for instance, a viable solution is $b_{ji}=b_i$,
$b_{ii}=-b_i$, $a_i=a$.  Further choosing RR fluxes 
\beq
q_i = -c_2 \ , \ \ e_i = c_1
\eeq
we allow a configuration with $T_1 = T_2 = T_3 = T$. Then the
superpotential (\ref{wa}) reduces to  
\beq 
W=e_0 + 3ic_1 T + 3c_2 T^2 + im T^3 +
ih_0 S - 3a ST - \sum_{k=0}^3 (ih_k + b_kT) U_k\ ,
\label{sup2}
\eeq 
If the fluxes $h_k$ and $b_k$ are independent of $k$,
we can also set $U_1=U_2=U_3=U$.
Given the fluxes leading to (\ref{sup2}), the tadpole conditions (\ref{lio1}) 
become \cite{vz, cfi} 
\beqa 
\sum_a N_a n_a^1 n_a^2 n_a^3 \ + \  \frac{1}{2}(h_0 m - 3 a c_2) & = & 16 \  ,
\nonumber \\
\sum_a N_a n_a^1 m_a^2 m_a^3 \ + \ \frac12(h_1 m + b_1 c_2) & = & 0  \  ,
\nonumber \\
\sum_a N_a m_a^1 n_a^2 m_a^3 \ + \ \frac12(h_2 m + b_2 c_2) & = & 0  \  ,
\nonumber\\
\sum_a N_a m_a^1 m_a^2 n_a^3 \  + \ \frac12(h_3 m + b_3 c_2) & = & 0 \ .
\label{tadodhiso} 
\eeqa 
We will examine the case of AdS vacua with negative cosmological constant 
in the non-supersymmetric models considered in the previous sections.
We recall that non-susy intersecting models without fluxes 
\cite{imr, kokos1, kokos2a, kokos2b} are unstable due to uncanceled 
NSNS tadpoles. However some of the tadpoles vanish as all the complex structure moduli could get fixed 
by the supersymmetry conditions present in the Pati-Salam models of \cite{kokos1}. 

The important observation to make is that the Pati-Salam models 
considered in the previous section can be also be considered as possible solutions to 
models that accommodate R/NS and metric fluxes. A similar observation has already 
been made in the appendix of \cite{cfi}, where it was shown that the N=0 models of \cite{imr}(that had no explicit susy on intersections)
could stabilize some moduli in the presence of metric fluxes.

We consider the case of isotropic K\"ahler moduli by 
choosing $T_k=T$ and look for minima of the effective supergravity potential 
with \beq
D_{U_k} W \ = \ 0 \ , \ D_S W \ = \ 0
\label{min}
\eeq
where $W$ given in
(\ref{sup2}) and as usual $D_{X} \ = \ \partial_X W + W \partial_K W$.
The above conditions imply the following constraint \cite{cfi} 
\beq
3a s = b_k u_k \ .
\label{fix}
\eeq
where $a$ and $b_k$ must be both non-zero and of the same sign.
Moreover, the following consistency conditions also hold
\beq
3h_k a + h_0 b_k = 0
\quad ; \quad k=1,2,3 \ .
\label{finetune}
\eeq
Before considering the issue of moduli stabilization let us mention one further 
constraint that should be imposed on the models.  
In general U(1) fields coupled to RR fields give rise to 
massive U(1)'s that are generated by the couplings
\beq F^a \ \wedge \ N_a\sum_{I=0}^3 \ c_I^a C_I^{(2)}
\label{bf}
\eeq
with
\beq c_0^a=m_a^1m_a^2m_a^3 \ ;\ c_1^a=m_a^1n_a^2n_a^3 \ ;\
c_2^a=n_a^1m_a^2n_a^3 \ ;\ c_3^a=n_a^1n_a^2m_a^3 \ \ .  
\eeq
In addition, metric backgrounds give rise - to avoid 
inconsistencies resulting from loss of gauge invariance - to the
constraint \cite{cfi}
\beq \int_{\Pi_a}\ (\ov{H}_3\ + \omega J_c) \ = \ 0 \  ,
\label{fwmetric}
\eeq
evaluated at the vacuum under examination, where  $\Pi_a$ denotes the 3-cycle
wrapped by the D6-brane, $\omega $ are the metric fluxes and
$J_c$ is the complexified K\"ahler 2-form of the torus.
Assuming a weaker form of this constraint that neglects metric fluxes, one gets
the weaker condition \cite{cfi}
\beq 
\sum_{I=0}^3\ c^a_I \ h_I \ =\ 0 \ \ .
\label{fwdir}
\eeq
The above condition guarantee that the U(1)'s that are getting 
massive from their couplings to RR fields are orthogonal to those one's becoming 
massive from NS \footnote{In general from NS and metric, see (\ref{fwmetric}).} 
fluxes.

\subsection{Moduli stabilization from fluxes}

Conditions (\ref{fix}), (\ref{finetune}), (\ref{fwdir}) may be used to examine the 
models of 
the previous sections in the presence
of general flux backgrounds. 
Imposing the condition (\ref{fwdir}) - on the general solution to the RR tadpoles 
(\ref{RR}) seen in table (\ref{skaB}) of appendix B - 
we derive for the left-right symmetric (the result is also valid for the Pati-Salam 
models and the MSSM coming from Wilson line breaking of the L-R)
\beq
h_2 \ = \ \frac{{\tilde \e}}{\e} \frac{\beta_2}{\beta_1} \ h_3 \ \ .
\label{con2}
\eeq
Observe the similarity  of this condition to the susy condition (\ref{susy}). 
In the present models 
that possess N=1 supersymmetry in the observable MSSM and a different $N^{\prime}=1$
supersymmetry in the supersymmetry breaking messenger sector, the hypercharge does survive 
massless the GS mechanism. It is given e.g. in the simplest left-right models as we have seen 
by eqn. (\ref{sda1}) or by (\ref{hypM}) when the breaking to the MSSM is achieved.
Notice that in the non-supersymmetric models with intersecting D6-branes
like \cite{imr} in the case that the RR tadpole parameters allow 
the hypercharge to  
survive massless to low energies [and thus having the SM chiral spectra at low energies
 ] {\em only the dilaton S may be fixed by the fluxes} 
as $h_1 = h_2 = h_3 = 0$ and $h_0 \neq 0$. Similar results are expected for the non-susy models of 
\cite{kokos2a,kokos2b} which are the maximal SM generalizations of \cite{imr} to five and
six stacks of intersecting D6-branes. The difference with \cite{imr}, is that in the latter models \cite{kokos2a,kokos2b}
we have in addition to the two extra U(1)'s which survive massless to low energies,   
one or two extra U(1)'s \footnote{For the 5-stck SM \cite{kokos2a} there are one of two U(1)'s; for the 6-stack SM \cite{kokos2b} there are only two U(1)'s.} 
which are expected to get broken by the vev of the 
sneutrino \footnote{The intersections that accommodate the right handed neutrino
in \cite{kokos2a,kokos2b} respect N=1 supersymmetry in sharp contrast with the \cite{imr} models
where the intersections and hence the chiral fermions preserve no susy at all.}. 
On the contrary in the present models - where N=1 susy    
makes its presense manifest - all real parts of moduli, S, T, U, namely 
s, t, $u_k$ are expected to be
fixed and also some linear axion combinations will get fixed.

From conditions (\ref{lio1}), (\ref{con2}), (\ref{susy}) we observe that
\beq
\frac{Re U_3}{Re U_2}  =  \frac{h_2}{h_3} =  \frac{{\tilde \e}}{\e} \frac{\beta_2}{\beta_1}
\label{la1}
\eeq
That means that {\em the supersymmetry conditions in the present models 
could fix the NS flux coefficients 
in the RR tadpoles}.  \newline
Hence if $\b_1 =  \b_2 = 1$, $h_2 \ = \ h_3$. From (\ref{la1}) we also 
conclude \footnote{For a two dimensional torus where the generating lattice
is spanned by the basis vectors $\vec{e}_i = G_{ij}e^j$; the complex structure moduli is given by $U \stackrel{def}{=}\frac{1}{G_{11}} (\sqrt{G} -i G_{12})$  },
 as we examined the case that there is no tilt in any of our tori,
 that 
\beq
Re U_2 \ = \ Re U_3 
\eeq
ans since our torus is orthogonal,
\beq
 \ \ \ Im U_1 = Im U_2 \ = \ Im U_3 = 0
\label{choi1}
\eeq
We are choosing a vacuum with $e_0 = c_1 = 0$, $h_0 \ = \ - 3 a $, $c_2 = - m$, 
\beq
\e \ = \ {\tilde \e}\ = \ \beta_1 \ = \ \beta_2 \ = \ 1
\eeq
and which satisfies the consistency conditions (\ref{finetune}) 
\beq
b_k \ = \ (6, 6, 6)\ , \ \ h_k \ = \ -\frac{h_0}{3a}\ b_k ,\ \ k \ = \ 1,2,3
\eeq
and where also $h_0 < 0$, $ a < 0$, $m < 0$.
The flux contributions to the RR tadpoles cancel as we have chosen
\beq
h_0 m \ = \ 3 a c_2 , \ \  \ h_i m \ + \ b_i c_2 \ = \ 0 \ ,
\label{cond12}
\eeq
e.g. the contribution of metric fluxes cancels against the 
contribution from R/NS fluxes. As this is a general result \cite{cfi}
it appears that : {\em One can choose any four dimensional 
 non 
supersymmetric vacuum on toroidal orientifolds which satisfies RR tadpoles
in the context of compactifications of type IIA theory with
intersecting D6-branes and make it to satisfy the RR tadpoles with fluxes} 
by making the choice (\ref{cond12}). \newline
The previous observation extends easily to the N=1 supersymmetric orientifold
vacua based on $Z_N$, $Z_N \times Z_M$ orbifolds as one has to add the contribution
 of the corresponding crosscap contributions to the RR tadpoles; one has to add
a contribution of $-16$ to the right hand side of the last three lines of 
(\ref{tadodhiso}). 

Let us here discuss some points regarding the fate of U(1)'s present. In general the U(1) combination \cite{cfi}
\beq
Im \ U_2 - \frac{{\tilde \e} \beta_2}{ \e    \beta_1 } Im \ U_3
\eeq 
gets massive from its coupling to RR BF fields (\ref{coup10}).
Its orthogonal combination 
\beqa
{\tilde X} \equiv Im \ U_2 + \frac{{\e} {\beta_1}}{{\tilde \e} \beta_2}\ Im \ 
U_3 & = &  
\frac{1}{h_2}(h_2 \ Im \ U_2 + h_3 \ Im \ U_3)
\eeqa
is expected to receive a mass from fluxes upon minimization as it should appear 
in the superpotential W.
Also massive becomes the combination of complex structure axions defined as \cite{cfi}
\beq
3 a Im S + \sum_k b_k Im U_k =  3 c_1 + \frac{3 c_2}{a}(3h_o - 7 a \upsilon) - 
\frac{3 m }{\upsilon}(3 h_0 - 8 a \upsilon), 
\label{mas1}
\eeq
where $\upsilon = Im T$  takes different values \footnote{In the following we assume $ m \neq 0$. }
according to whether $m =0$, 
 or $m \neq 0$.
For the vacua associated to (\ref{choi1}), this combination becomes
   $3 a Im S + b_1 Im U_1$. Thus given that $Im U_i$ are fixed from (\ref{choi1}), 
the ImS is 
fixed in (\ref{mas1}). 
In general - neglecting NS tadpole contributions - the real part of the S, $T^i$, $U^i$ moduli denoted as $s, u_k, t$ respectively get also fixed \footnote{Borrowing the results from \cite{cfi}.} in the present models  
at   
\beq
s = \frac{m}{a 10^{1/3}}\frac{- h_0}{3 m} \ t , \ \ \ u_k = \frac{3 a s}{b_k}, \ \ 
t = \frac{ \sqrt{15} 10^{2/3} }{20} \ \frac{- h_0}{3 m}
\eeq
leaving unfixed only the imaginary parts of the K\"ahler moduli.
The latter could be fixed when the NS potential is included.
For some other proposals to fix the K\"ahler moduli see \cite{bitre, bika}.
In order to trust the supergravity approximation that we are using
it is necessary that $-h_0/3a$ large since the 4-D dilaton 
$e^{\phi_4} = (s u_1 u_2 u_3)^{1/4}$.

We also note that that as the present models have uncanceled NSNS tadpoles,
the exact stabilization problem is solved by considering the 
minimization of the full potential that includes the NSNS potential in addition
to the potential generated by the superpotential fluxes. For simplicity reasons we will not 
consider this option here leaving the full treatment in future work.


\section{Conclusions}

We have presented the first three generation D-brane string models that localize the 
MSSM spectrum at the string scale, 
simultaneously providing us with a particle physics model that accommodates all
the long seeking properties of a realistic MMSM incorporating string vacuum, that possess : 
\begin{itemize}
\item Unification at $2 \cdot 10^{16}$ GeV
(of SU(3) and SU(2) gauge couplings).  
\item All extra chiral and non-chiral beyond the MSSM matter is getting massive at the string scale.
\item Proton stability, as baryon number is a gauged symmetry.
\item The MSSM spectrum is massless at the string scale while the theory 
breaks to the SM $SU(3)_c \times SU(2) \times U(1)_Y$ at low energies.
\end{itemize}
The great advantage of using intersecting D-branes against other approaches involving
compactifications of the perturbative string theories 
is that in the present formalism we can guarantee proton stability, moduli fixing (through supersymmetry conditions at intersections) and exotics becoming massive 
occuring simultaneously.\newline
A comment is in order. Most of the non-chiral states become massive by 
the introduction of Scherk-Schwarz breaking. For the rest of them, including also
the chiral exotics of the vector-like messenger sector, trilinear superpotential
couplings (TSC) are employed, where it is assumed that the coupled gauge singlet directions do receive a vev. The current models provide us with a laboratory that 
can be used 
to investigate in the future the precise determination of these vevs.
We note that the presence of these TSC -- in the presence of a stable proton -- 
for the extra beyond the MSSM matter multiplets is absent in all 
the models from intersecting branes or other recent model building constructions 
from string theory \cite{cve,cve1,bluo1,ov,ott1,she,nano1,alda,ana,ms, blumen1}.
\newline
In toroidal orientifold vacua with intersecting D6-branes, supersymmetry
is already broken by construction as the absence of a variety of orientifold planes does not allow N=1 models but only N=0 models to be constructed. 
Thus the vacua presented in this work, even though they localize the MSSM multiplet spectrum among the observable sector D6-branes 
$O_i$, are 
non-supersymmetric (N=0); the N=1 supersymmetry of the extra messenger sector
is different from the one respected by the MSSM multiplets.

The most important future of the present string models is that they manage
to {\em make} all the extra chiral/non-chiral beyond the MSSM exotics 
{\em massive}, due to the combined use of the introduction of Scherk-Schwarz
deformations and flat directions. \newline
These models have the novel property that only the  
supersymmetric intersections share a susy 
with the orientifold plane and not the   
participating intersecting branes, meaning that some of the branes involved can 
be non-supersymmetric.

The Pati-Salam/left-right symmetric models of this work are the first realistic examples 
of gauge mediation in string theory. They provide us with a stringy platform for 
realistic calculations in the context of string theory. 
Attempts to construct gauge mediated 
semirealistic models in string theory could be 
found in \cite{hypo, anto6}, while local models of gauge mediation inspired from
string theory could be found in \cite{medi3}. In our models we don't have to show dynamical supersymmetry breaking in the standard sense as supersymmetry is broken by construction. 
The great advantage of gauge mediation 
against other schemes of supersymmetry breaking --- like gravity mediation or 
anomaly mediation --- are the universal 
squark and slepton (soft term) masses.
In the present constructions masses for sparticles which are due to the
D-term breaking and originate from FI terms in the limit of small deviations of complex structure result in degenerate squarks and slepton masses 
and thus may be  
responsible for automatic flavor changing neutral current suppression.

We also explained the "embedding" of the present models in toroidal orientifold 
 backgrounds of - type IIA 4D compactifications - in the  presence of 
NS/R and metric fluxes in the spirit of \cite{dkpz}, \cite{cfi} (and \cite{vz}).
The stringy toroidal orientifolds of sections 2,3,4,5 that in the absence of fluxes have NS tadpoles remaining,
now become stable, as the real and imaginary parts of dilaton and 
complex structure moduli and also the real parts of K\"ahler moduli get 
fixed.
The only remaining unfixed moduli, the imaginary parts of K\"ahler ones, could be fixed by the complete minimization of the NS scalar potential
\beq
V = \frac{T_6}{\lambda}\sum_a (||l_a||-||l_{o}||),
\eeq
where $T_6$ the tension of the D6-branes; $\lambda$ the string coupling;
$||l_a||, ||l_{o}||$ 
the $T^6$ volume that the D-branes and the O6-planes lie respectively.
The uplifting of Ads vacua to de Sitter ones is an interesting possibility once the NS potential
is introduced, without the need of any non-perturbative effects \cite{kah1}, and it will be examined elsewhere.\newline
An important comment is in order. When the non-supersymmetric models of \cite{imr} were 
examined in the presence of RR/NS and metric fluxes \cite{cfi} - in these models there is no supersymmetry present at any intersection - when
the hypercharge survives massless to low energies \cite{kokos2a,kokos2b}, only the dilaton could be fixed by fluxes. On the contrary in the
present non-supersymmetric models where N=1, N=2 (Higgses/higgsinos) supersymmetry makes explicit its appearance at intersections, the hypercharge survives massless and both the dilaton
and the complex structure moduli could be fixed. 
Also in the present constructions more moduli are fixed 
than in the usual N=1 Ads vacua of \cite{cfi} - where only the real parts of the dilaton and complex 
and K\"ahler moduli get fixed - as the toroidal 
lattice fixes also the imaginary parts of the complex structure moduli.

It will be also interesting to examine moduli stabilization in the present
models in the spirit of \cite{inanew}. 
We note that when passing from the stringy MSSM models of sections 3, 4, 5
to the fluxed models of section 6, the only feature of the string models 
that is not surviving in the presence of fluxes is the non-chiral matter that is getting massive
 by the introduction of SS breaking.

We also note that the universal value obtained for the sparticle 
masses coming from D-term breaking depend on the small variation of the
complex structure and thus on the tori radii. For particular values 
of these radii, the scale of sparticle masses could be lowered 
well below the string GUT/unification scale of $2.04 \times 10^{16}$ GeV where the $SU(3), SU(2)$
gauge couplings meet.

Our work also confirms that proposals based on the idea of statistical analysis \cite{do} of the 
existence of 
a string landscape \cite{su, do} 
which implicitly accepts the 
existence of a stringy Standard model vacuum with no exotics are definitely real.
Further studies of the present models including
studies of susy soft breaking terms and low energy squark/slepton phenomenology, 
quark and lepton mass hierarchies \cite{duno,flav1} may be examined elsewhere.



\begin{center}
{\bf Acknowledgments}
\end{center}

We would like to thank I. Antoniadis, C. Angelantonj, for discussions;
C.K would like to thank P. Camara, G. Giudice, S. Dimopoulos for a discussion.
C.K would like also to thank the Organizers of the String Phenomenology 2005, 
SUSY 2005,
HEP2006 conferences where the 
results of this work were announced and in addition the Cern theory
division for their warm hospitality where parts of this work were completed.
E.G.F and C.K are supported by the research program 
Pithagoras II, grant no : 70-03-7992 of the Greek Ministry of National 
Education that is 
also funded by 
the European Union. 


\section{Appendix A -Alternative Hypercharge embedding for the L-R symmetric models}

There is alternative hypercharge embedding for the 
left-right symmetric models discussed in section (3.2) that gives the same
spectrum and also the same
hypercharge for the observable MSSM related part of the spectrum in table 
(\ref{taba1}).
It is defined as 
\beq
U(1)_{\tilde Y} = \frac{1}{6} Q_a - \frac{1}{2} Q_d -\frac{1}{2} Q_{h_3} -
\frac{1}{2} Q_{h_4}
\label{assi1}
\eeq 
Notice that the hypercharge assignment of the messenger multiplets are different than the one 
appeared assigned to the choice of eqn. (\ref{sda1})
on table (\ref{taba1}). The RR couplings reveal that the U(1) 
defined as $3 Q_a + Q_d$, 
gets massive while the extra U(1)'s $U(1)^{(1*)}= Q_{h_3} - Q_{h_4}$, $U(1)^{(2*)}= 
(6 Q_a - 18 Q_d) + 10(Q_{h_3} + Q_{h_4})$ get broken if ${\ti \phi}_{R}^1$, ${\ti \phi}_{L}^3$ get vevs respectively as they are flat 
directions of the potential.
Also non-chiral matter gets massive by SS deformations as in section (3.2).
However the couplings that are available for the 
messenger multiplets within the new assignment (\ref{assi1}) 
are exactly the same that appear in eqns (\ref{coup1}).
As a result we obtain for the different hypercharge embedding (\ref{assi1}) 
the same physics; same unification scale, all exotics massive, etc.


\section{Appendix B}

We this appendix we provide the solution to the RR tadpoles for the 
Pati-Salam models of section  3, in the case that 
\beq
\e = {\ti \e} \ , \ \rho=1,1/3 \ .
\eeq
The case $\rho =1/3$ was treated in table 6 and is used to facilitate the discussion in
the main text of this work. These solutions appear in table (\ref{skaB}). 


\begin{table}
[htb]\footnotesize
\renewcommand{\arraystretch}{2.3}
\begin{center}
\begin{tabular}{||c||c|c|c||c|||||} 
\hline
\hline
$\ N_i$ & $(n_i^1, \ m_i^1)$ & $(n_i^2, \ m_i^2)$ & $(n_i^3, \ m_i^3)$& G.G. 
$->$ $\epsilon =+1$  \\
\hline\hline
 $N_a=4$ & $(1,\ 0)$  & $(\frac{1}{\rho}, \  3 \rho \e \beta_1)$ & $(\frac{1}{\rho}, 
\  - 3 \rho {\e} \beta_2)$ 
& $U(4)$  \\
\hline
$N_b=1$  & $(0, \ 1)$ & $(1/\beta_1,\ 0)$ & $(0, \ -{\e})$ & $Sp(2)$  \\
\hline
$N_c=1$ & $(0, \ \e)$ & $(0,\  -\e)$  & $({\e}/\beta_2, \ 0)$ & $Sp(2)$ \\    
\hline\hline
$N_{h^1} = \frac{4}{\rho^2}\beta_1 \beta_2$ & $(1,\ 0)$  &
$(-1/\beta_1, \  0)$ & $(1/\beta_2, \ 0)$&$U(1)^{N_{h_1}}$  
 \\\hline
$N_{h^2}= 36 \rho^2 \beta_1 \beta_2 $  & $(1, \ 0)$ & $(0,\ {\e})$ & $(0, \ {\e})$ & $U(1)^{N_{h_2}}$ \\\hline
$N_{h^3}=1$ & $(0, \ 1)$ & $(1/\beta_1,\  0)$  & $(0, \ {\e})$ & $U(1)$ \\    
\hline
$N_{h^4}=1$ & $(0,\ {\epsilon})$ &  $(0,\  \epsilon )$  
  & $({\epsilon}/\b_2, \ 0)$ & $U(1)$  \\\hline
$N_{h^5}=16\beta_1 \beta_2$ & $(1,\ 0)$ &  $(1/\beta_1,\ 0 )$  
  & $(1/\beta_2, \ 0)$ 
& $Sp(2)^{N_{h_5}}$\\\hline\hline
\end{tabular}
\end{center}
\caption{\small 
Solution to the RR tadpoles for toroidal 
orientifold models. The N=1 MSSM chiral spectrum arises as part of Pati-Salam models 
$SU(4) \t Sp(2)_b \t Sp(2)_c$ in the top part of the table from intersections 
between a, b, c, d branes.  
Messenger multiplet states respecting a $N^{\prime}=1$ supersymmetry 
arise from the intersections of the a, d branes with the 
$h_3$, $h_4$ branes. 
We have set $\epsilon = {\ti \epsilon}$ in table (\ref{spe1}); the parameter 
$\rho$ can take the values $1, 1/3$.
\label{skaB}}          
\end{table}







%


\begin{thebibliography}{900}


\bibitem{cve}
M. Cvetic, G. Shiu,
A. M. Uranga,  
 Nucl. Phys. B615 (2001) 3, hep-th/0107166;
M. Cvetic, G. Shiu, A. M. Uranga, Phys. Rev. Lett. 87 (2001) 20,
hep-th/0107143


\bibitem{cve1}M. Cvetic, T. Li, T. Liu, 
Nucl.Phys. B698 (2004) 163, hep-th/0403061;\\
M. Cvetic, P. Langacker,  T. Li, T. Liu, 
Nucl.Phys. B709 (2005)  241, hep-th/0407178

\bibitem{lustr}R. Blumenhagen, B. Kors, D. Lust, {\em Type I Strings with F- and B-Flux}, JHEP 0102 (2001) 030
hep-th/0012156.

\bibitem{review1} R. Blumenhagen, M. Cvetic, P. Langacker, G. Shiu, 
Ann.Rev.Nucl.Part.Sci. 55 (2005) 71, hep-th/0502005.

\bibitem{ampli}S.Abel, A.W.Owen, Nucl.Phys. B663 (2003)197 
hep-th/0303124; M.Cvetic and I.Papadimitriou,
Phys.Rev.D68:046001,2003, Erratum-ibid.D70 (2004) 029903, 
hep-th/0303083


\bibitem{review2}D.Lust, Class.Quant.Grav.21:S1399 (2004),  hep-th/0401156


\bibitem{kokore}C. Kokorelis, {\em Standard Model Building from Intersecting D-Branes},
Appeared in the series {\em New developments in String Theory Research}, Nova Publishers, NY , 
2005, hep-th/0410134



\bibitem{bluo1}
R. Blumenhagen, L. G\"orlich, T. Ott, 
JHEP 0301 (2003) 021,  hep-th/0211059;
 R. Blumenhagen, J. P. Conlon, K. Suruliz, 
JHEP 0407:022,2004, hep-th/0404254 



\bibitem{ov} V. Braun, Y.-H. He, B. A. Ovrut, T. Pantev, 
hep-th/0512177

\bibitem{ov1}T.L. Gomez, S. Lukic, I. Sols, {\em Constraining the Kahler moduli in the 
Heterotic Standard Model}, 
hep-th/0512205


\bibitem{sa}C. Angelantonj and A. Sagnotti, 
arXiv:hep-th/0010279 


\bibitem{bi}M. Larosa and G. Pradisi, 
Nucl. Phys. B667:261, 2003, hep-th/0305224

\bibitem{bitre}M.Bianchi and E.Trevigne, JHEP 0508, 034 (2005), hep-th/0502147 


\bibitem{imr}
L.~E.~Ib\'a\~nez, F.~Marchesano and R.~Rabad\'an,
{\em Getting just the standard model at intersecting branes},
JHEP, 0111 (2001) 002, hep-th/0105155



\bibitem{kokos1} C. Kokorelis,JHEP 08 (2002) 018, hep-th/0203187;
{\em GUT model Hierarchies from Intersecting Branes}, 
C.Kokorelis, hep-th/0210004;  C. Kokorelis, JHEP 0211 (2002) 027,   hep-th/0209202]; C. Kokorelis, {\em Deformed Intersecting D6-Branes II}, hep-th/0210200


\bibitem{kokos2a}C. Kokorelis,
{\em New Standard Model Vacua from Intersecting Branes},
JHEP 09 (2002) 029,
hep-th/0205147

\bibitem{kokos2b}C. Kokorelis,
{\em Exact Standard Model Compactifications from Intersecting Branes},
JHEP 08 (2002) 036, hep-th/0206108

\bibitem{z3}R. Blumenhagen, B. Kors, D. Lust, T. Ott, Nucl. Phys. B616 (2001) 3, 
hep-th/0107138


\bibitem{Iba}D.Cremades, L.Ibanez , F.Marchesano, JHEP 0207 (2002) 009, hep-th/0201205

\bibitem{cimyuks}
D.Cremades, L.Ibanez, F.Marchesano, {\em Computing Yukawa couplings from magnetized 
extra dimensions}, JHEP {0405} (2004) 079, hep-th/0404229.
 


\bibitem{ott1}G.Honecker and T.Ott, Phys.Rev.D70:126010,2004, Erratum-ibid.D71:069902,2005,
hep-th/0404055

\bibitem{she}T.P.T. Dijkstra, L. R. Huiszoon and A. N. Schellekens,
hep-th/0403196


\bibitem{bfield}M.Bianchi, G.Pradisi and A.Sagnotti, Nucl. Phys. B376 (1992) 365;
C.Angelantonj, Nucl. Phys. B566 (2000) 126; Z.Kakushadze, Int.J.Mod.phys.A15 (2000) 3113 



\bibitem{cre}D. Cremades, L.E. Ibanez, F. Marchesano, 
{\em Towards a Theory of Quark Masses, Mixings And CP Violation}, hep-ph/0212064 

\bibitem{lo1}Cremades, Ibanez, Marchesano, JHEP 0307 (2003) 038,  
hep-th/0302105

\bibitem{kokosusy}C. Kokorelis, {\em N=1 Locally Supersymmetric 
Standard Models from Intersecting Branes}", 
 hep-th/0309070, Revised version  to appear 

\bibitem{quasi}D. Cremades, L.E. Ibanez, F. Marchesano,
JHEP 0207 (2002) 022, hep-th/0203160



\bibitem{klewi}I.Klebanov and E. Witten, Nucl. Phys. B664 (2003) 3, 
hep-th/0304079


\bibitem{afk}M.Axenides, E.Floratos and C.Kokorelis, JHEP0310 (2003) 006,
hep-th/0307255 

\bibitem{bur1}P. Burikham, JHEP 0502 (2005) 030, hep-ph/0502102

\bibitem{nathp}P. Nath, P. Fileviez Perez,  hep-ph/0601023;
R.Tatar, T.Watari, hep-th/0602238

\bibitem{cr1}M. Cvetic, R. Richter, hep-th/0606001

\bibitem{bere} D. Berenstein, hep-th/0603103

\bibitem{lebe} W. Buchmuller, K. Hamaguchi, O. Lebedev, 
M. Ratz, hep-th/0606187


\bibitem{gm}S. Dimopoulos, S. Raby, Nucl. Phys. B192 (1982) 353;
M. Dine, W. Fischler, M. Srednicki, Nucl. Phys. B189 (191) 575;
M.Dine, W.Fischler, Phys. Lett. B110 (1982) 227;
Nucl. Phys. B204 (1982) 346;C. Nappi, B.Ovrut, Phys. Lett. B113 (1982) 175;
L. Alvarez-Gaume, M. Claudson and  B. Wise, Nucl. Phys. B207 (1992) 96;
S. Dimopoulos, S. Raby, Nucl. Phys. B219 (1993) 479;
M.Dine, A.E.Nelson, Phs. Rev. D48 (1993) 1277;
D51 (1995) 1362;M.Dine, A.E.Nelson, Y.Shirman, Phys. Rev. D51 (1995) 1362;
M.Dine and A.E.Nelson, Y.Nir, Y.Shirman, Phys. Rev. D53 (1996) 2658;
G.F.Guidice and R.Ratazzi, Phys. Rept. 322 (1999) 419;
S.L.Dubovsky, D.S.Gorbunov, S.V.Troitsky, Phys.Usp.42 (1999) 623;
S.Dimopoulos, S.Thomas, J.D.Wells, Nucl. Phys. B488 (1997) 39 


\bibitem{blu}Blumenhagen, Lust, Stieberger, "Gauge Unification in supersymmetric intersecting
brane worlds", JHEP 0307 (2003) 036,  hep-th/0305146


\bibitem{kane2}G. L. Kane, P. Kumar, J. D. Lykken, T. T. Wang, 
Phys.Rev.D71 (2005) 115017, hep-ph/0411125



\bibitem{blt}R.Blumenhagen, D.Lust, T.Taylor,  Nucl.Phys. B663 (2003) 319, hep-th/0303016

\bibitem{cu}J.F.G.Cascales, A.M. Uranga,  JHEP 0305 (2003) 011, hep-th/0303024;
hep-th/0311250

\bibitem{ms}F.Marchesano, G.Shiu, "Building MSSM flux vacua", 
JHEP 0411 (2004) 041,  
hep-th/0409132 

\bibitem{DeWolfe}
O.~DeWolfe, A.~Giryavets, S.~Kachru and W.~Taylor,
{\em Type IIA moduli stabilization}, hep-th/0505160.


\bibitem{dkpz} J.~P.~Derendinger, C.~Kounnas, P.~M.~Petropoulos and
F.~Zwirner, 
Nucl.\ Phys.\ B715 (2005) 211, hep-th/0411276; 
hep-th/0503229.

\bibitem{vz} G. Villadoro, F. Zwirner, JHEP 0506 (2005) 047, hep-th/0503169


\bibitem{cfi}P.G. C\'amara, A. Font,  L.E. Ib\'a\~nez, JHEP 0509 (2005) 013,
hep-th/0506066



\bibitem{SSta1}I. Antoniadis and T. Taylor, Nucl. Phys. B695 (2004)103,
hep-th/0403293


\bibitem{split0}N.Arkani-Hamed and S. Dimopoulos, JHEP 0506:073,2005,
  hep-th/0405159


\bibitem{split}N. Arkani-Hamed, S. Dimopoulos, G.F. Giudice, 
A. Romanino, "Aspects of Split Supersymmetry",
Nucl.Phys. B709 (2005) 3, hep-ph/0409232



\bibitem{split1}I. Antoniadis and S. Dimopoulos,  {\em Split supersymmetry in string 
theory},  Nucl.Phys.B715 (2005) 120, 
 hep-th/0411032 

\bibitem{split2}C.Kokorelis, {\em Standard Models and Split Supersymmetry
from Intersecting Brane Orbifolds}, 
 hep-th/0406258

\bibitem{orbiko}C.Kokorelis, {\em Standard Model Compactifications of IIA Z3 x Z3 Orientifolds from Intersecting D6-branes}, 
 Nucl.Phys. B732 (2006) 341,  hep-th/0412035


\bibitem{split3}B. Kors and P. Nath, hep-th/0411201


\bibitem{split4}N.Haba, N. Okada, hep-ph/0602013;
S. Kumar Gupta, B. Mukhopadhyaya, S. Kumar Rai, hep-ph/0510306;
  E. Dudas, S. K. Vempati, hep-th/0506172; A.Ibarra, hep-ph/0503160;
N.G. Deshpande, J. Jiang, hep-ph/0503116; B. Dutta, Y. Mimura, hep-ph/0503052;
N. Haba, N. Okada, hep-ph/0502213]; G. Senjanovic, hep-ph/0501244;
 Z. Lalak, R. Matyszkiewicz, hep-ph/0506223;
K.S. Babu, Ts. Enkhbat, B. Mukhopadhyaya, hep-ph/0501079;
S. P. Martin, K. Tobe, J. D. Wells,hep-ph/0412424; P. Fileviez Perez, hep-ph/0412347;
P. C. Schuster, hep-ph/0412263;
A. Datta and X. Zhang, hep-ph/0412255; A. Masiero, S. Profumo, P. Ullio, hep-ph/0412058


\bibitem{ps}J.Pati and A.Salam, Phys.Rev.D10, 275 ,1974;
see also J. Pati,  hep-ph/0407220 and  
J.Pati, {\em Probing Grand Unification Through Neutrino Oscillations, Leptogenesis, 
and Proton Decay}, hep-ph/0305221

\bibitem{D51}
C. ~Kokorelis, 
{\em Exact Standard model Structures from Intersecting D5-branes}, Nucl. 
Phys. B677:115, 2004,  
hep-th/0207234

\bibitem{D52}D. Cremades, L.~E.~Ib\'a\~nez and
 F.~Marchesano,
 {\em Standard model at intersecting  D5-Branes: lowering the string scale},
 Nucl. Phys. B643 (2002) 93, hep-th/0205074


\bibitem{leo} 
A. Prikas, N.D. Tracas, hep-ph/0303258;
D.V. Gioutsos, G.K. Leontaris, J. Rizos, Eur.Phys.J. C45:241(2006),
hep-ph/0508120;D.V. Gioutsos, hep-ph/0605278


\bibitem{alda}G. Aldazabal, E. Andres, J. E. Juknevich, 
{\em On Susy Standard-like models from orbifolds of D=6 Gepner orientifolds},
hep-th/0603217

\bibitem{hone}R. Blumenhagen, G. Honecker, T. Weigand, hep-th/0510050

\bibitem{ana}P. Anastasopoulos, T.P.T. Dijkstra, E. Kiritsis, A.N. Schellekens, hep-th/0605226


\bibitem{cw}J. Kumar and J.D. Wells, JHEP 0509, 067 (2005), hep-th/0506252; hep-th/0604203; J.Kumar, hep-th/0601053


\bibitem{ai}
C.~Angelantonj, M.~Cardella and N.~Irges, {\em Scherk-Schwarz breaking and intersecting branes}, Nucl. Phys. B725 (2005) 115  
hep-th/0503179.

\bibitem{openmoduli}
P.~G. C\'amara, L.~E. Ib\'a\~nez, and A.~M. Uranga,
{\em Flux-induced SUSY-breaking soft terms}, 
Nucl.\ Phys.\ B689, 195 (2004), hep-th/0311241; 
M.~Gra\~na, T.~W. Grimm, H.~Jockers, and J.~Louis, 
Nucl. Phys. B690 (2004) 21, hep-th/0312232; 
L.~G\"orlich, S.~Kachru, P.~K.~Tripathy and S.~P.~Trivedi,
hep-th/0407130; D.~L\"ust, P.~Mayr, S.~Reffert and S.~Stieberger,
hep-th/0501139.


\bibitem{duti}E.Dudas, C.Timirgaziu, {\em Internal magnetic fields and 
supersymmetry in orientifolds}, 
Nucl. Phys. B716 (2005)65, hep-th/0502085  

\bibitem{ver}H.Verlinde and M.Wijhholt, {\em Building the SM on a D3-brane}, 
hep-th/0508089

 
\bibitem{cve10}M. Cvetic, T. Li, T. Liu
{\em Standard-like Models as Type IIB Flux Vacua}, Phys.Rev. D71 (2005) 106008,
hep-th/0501041


\bibitem{nano1} C-M. Chen, T.Li, D.V.Nanopoulos, Nucl. Phys. B732 (2006) 224,
hep-th/0509059; C-M. Chen, V.E.Mayes, D.V.Nanopoulos, Phys.Lett. B633 (2006) 618, hep-th/0511135


\bibitem{duno}B.Dutta and Y.Mimura, {\em Properties of fermion mixing in intersecting 
D-brane models}, Phys.Lett. B633 (2006) 761, hep-ph/0512171;
hep-ph/0604126


\bibitem{ki}F.Wang, W.Wang, J. Yang, hep-ph/0512133 

\bibitem{dila}L. Ibanez, C. Munoz and S. Rigolin, Nucl.Phys. B422 (1994) 125, 
hep-ph/9308271 

\bibitem{babu1}K. S. Babu, B. Dutta, R. N. Mohapatra, "Up-Down Unification, Neutrino Masses and Rare Lepton Decays",
Phys.Lett. B458 (1999) 93, hep-ph/9904366 
 
\bibitem{flav1}T. Higaki, N. Kitazawa, T. Kobayashi, K. Takahashi, Phys.Rev. D72 (2005) 086003, hep-th/0504019

\bibitem{hypo}D. Diaconescu, B. Florea, S. Kachru, P. Svrcek, JHEP 0602 (2006) 020, 
hep-th/0512170. 

\bibitem{anto6}I.Antoniadis, K.Benakli, A.Delgado, M.Quiros, M.Tuckmantel, 
hep-th/0601003

\bibitem{popu}C. Munoz, Int. J. Mod. Phys. A19 (2004)3093 ,
hep-ph/0309346;
G. Bertone, D. Hooper, J. Silk,
Phys. Rept. 405(2004) 279,  hep-ph/0404175


\bibitem{smith}D.R.Smith and N.Weiner, Phys. Rev. D64 (2001)043502, 
hep-ph/0101138


\bibitem{inanew}G. Aldazabal, P.G. Camara, A. Font, L.E. Ibanez,
"More Dual Fluxes and Moduli Fixing", hep-th/0602089

\bibitem{medi3}
 D. Berenstein, C.P. Herzog, P. Ouyang, S. Pinansky,
JHEP 0509:084,2005, 
hep-th/0505029;
K. Intriligator, N. Seiberg, 
JHEP 0602:031,2006, hep-th/0512347;
I. Garcia-Etxebarria, F. Saad, A. M. Uranga,
hep-th/0605166


\bibitem{kah1}S.Kachru, R.Kallosh, A.Linde, S.P.Trivedi, {\em De Sitter vacua in string theory}, hep-th/0301240  

\bibitem{kah2}I.Antoniadis, A.Kumar, T.Mailard, {\em Moduli stabilization with open and closed string fluxes}, hep-th/0505260; \\
F.Denef, M.Douglas, B.Florea, A.Grassi, 
S.Kachru, Fixing all moduli in a simple f-theory compactification,
hep-th/0503124;\\
P.Berglund, P.Mayr, {\em Non-perturbative superpotentials in F-theory and string duality},
hep-th/0504058;\\
J.Conlon, F.Quevedo, K.Suruliz, {\em Large-volume flux compactifications: Moduli spectrum and D3/D7 soft supersymmetry breaking}, 
hep-th/0505076



\bibitem{blumen1} R. Blumenhagen, S. Moster, T. Weigand, hep-th/0603015


\bibitem{dimoka}N. Arkani-Hamed, S. Dimopoulos, S. Kachru
{\em Predictive Landscapes and New Physics at a TeV},
hep-th/0501082


\bibitem{du} M. R. Douglas, W. Taylor, {\em The landscape of intersecting brane
  models}, hep-th/0606109.


\bibitem{su}L.Susskind, {\em The anthropic landscape of string theory}, 
hep-th/0302219

\bibitem{do}M.R.Douglas, {\em The statistics of string M-theory vacua}, JHEP 0305, 046 (2003), hep-th/0303194;
S.Ashok, M.R.Douglas, {\em Counting flux vacua}, JHEP 0401, 060 (2004), hep-th/0307049;
F.Denef,  M.R.Douglas, {\em Distributions of string vacua}, JHEP 0405, 072 (2004), hep-th/0404116





\end{thebibliography}
\end{document}